\documentclass[12pt]{article}

\usepackage{amssymb,amsmath,amsfonts,eurosym,geometry,ulem,graphicx,caption,color,setspace,sectsty,comment,footmisc,pdflscape,subcaption,array,hyperref,placeins,afterpage,parskip,siunitx,physics,dcolumn,xcolor}
\edef\svtheparindent{\the\parindent}
\parindent=\svtheparindent\relax
\usepackage[utf8]{inputenc} 

\usepackage[
backend=biber,
style=apa,
uniquename=false,
uniquelist=false,
giveninits=true]{biblatex} 

\usepackage{csquotes}
\usepackage[american]{babel}
\DeclareLanguageMapping{american}{american-apa}

\DeclareMathOperator{\plim}{plim}
\graphicspath{ {./plots/} }

\newcolumntype{L}[1]{>{\raggedright\let\newline\\arraybackslash\hspace{0pt}}m{#1}}
\newcolumntype{C}[1]{>{\centering\let\newline\\arraybackslash\hspace{0pt}}m{#1}}
\newcolumntype{R}[1]{>{\raggedleft\let\newline\\arraybackslash\hspace{0pt}}m{#1}}

\numberwithin{equation}{subsection}
\begin{document}

\begin{titlepage}
\title{Understanding jumps in high frequency digital asset markets \thanks{This research was supported by the Deutsche Forschungsgesellschaft through the International Research Training Group 1792 "High Dimensional Nonstationary Time Series". This research is based upon work from COST Action 19130, supported by COST (European Cooperation in Science and Technology).}}
\author{Danial Saef\thanks{Humboldt-Universität zu Berlin, International Research Training Group 1792, Spandauer Str. 1, 10178 Berlin, Germany. PricewaterhouseCoopers, Germany. Email: danial.saef@pwc.com} \and
        Odett Nagy \thanks{Corvinus University of Budapest, Hungary. Email: odett.nagy@stud.uni-corvinus.hu}
        \and
        Sergej Sizov\thanks{Thieme Verlag, Germany. Email: sergej.sizov@thieme.de}   \and
        Wolfgang Karl Härdle\thanks{Humboldt-Universität zu Berlin, BRC Blockchain Research Center, Berlin; Sim Kee Boon Institute, Singapore Management University, Singapore; WISE Wang Yanan Institute for Studies in Economics, Xiamen University, Xiamen, China; Yushan Scholar National Yang-Ming Chiao Tung University, Dept Information Science and Finance, Hsinchu, Taiwan, ROC; Charles University, Dept Mathematics and Physics, Prague, Czech Republic; Grant CAS: XDA 23020303 and DFG IRTG 1792 gratefully acknowledged. Email: haerdle@hu-berlin.de}}
\date{\today}
\maketitle
\begin{abstract}
\noindent 
While attention is a predictor for digital asset prices, and jumps in Bitcoin prices are well-known, we know little about its alternatives. Studying high frequency crypto data gives us the unique possibility to confirm that cross market digital asset returns are driven by high frequency jumps clustered around black swan events, resembling volatility and trading volume seasonalities. Regressions show that intra-day jumps significantly influence end of day returns in size and direction. This provides fundamental research for crypto option pricing models. However, we need better econometric methods for capturing the specific market microstructure of cryptos. All calculations are reproducible via the \protect \includegraphics[height=0.5cm]{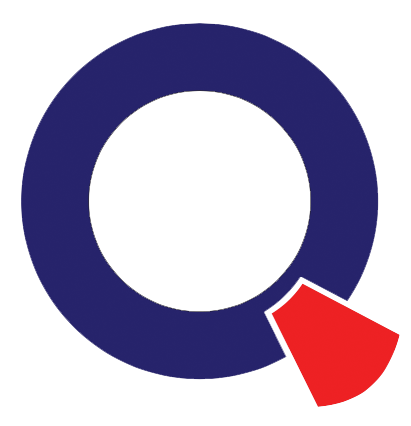} {\color{blue}\href{https://github.com/QuantLet/JumpDetectR}{quantlet.com}} technology.

\vspace{0in}


\bigskip
\end{abstract}
\setcounter{page}{0}
\thispagestyle{empty}
\end{titlepage}
\pagebreak \newpage

\doublespacing
\setlength\parindent{24pt}

\section{Introduction} \label{sec:introduction}
It is clear for sceptics that cryptos are a Ponzi scheme, and even clearer for supporters that cryptos are the future. But scientific knowledge about this new asset class is barely established. \textcite{liu_risks_2020} have identified predictors for cryptocurrency (CC) returns, such as investor attention and an affinity for extreme returns of +-5\% and larger, but have focused on daily data. Similarly, \textcite{aste_cryptocurrency_2019} finds that end of day prices are significantly correlated with sentiment, but studies only positive sentiment. \textcite{scaillet_high-frequency_2020} established that Bitcoin is subject to price jumps, but uses a dataset from the closed exchange MtGox observing the market from 2011-2013. \textcite{ohara_high_2015} has pointed out that changes in market microstructure due to high frequency (HF) and algorithmic trading require that empirical analyses account for the inter-connectedness of global markets due to cross-market arbitraging, and for the changes in time dimensions with respect to sampling frequency. To address these important gaps, we study the cross section of Bitcoin and its alternatives in different geographic regions and detect jumps at the highest possible frequency. This reveals that black swan events, be it from influencers in the CC space, regulators, or governments are a feature of every large cryptocurrency. They often trigger significant price jumps that spread through the CC network and can even affect currencies unrelated to that specific event.

We analyse a new dataset that is readily available for researchers. It contains tick data of the biggest CCs at the biggest exchanges and captures events like the Covid crisis and the early stage of the latest bull run. As there have long been discussions whether to include a jump component to option pricing models, such as in \textcite{duffie_transform_2000}, we make use of the HF econometrics toolbox and show that jumps are present in up to 65\% of observed trading days. Thus, we establish that jumps are an essential component in the price process of CCs. Our analysis reveals seasonality patterns both intra-daily and weekly that are similar to those in volatility and trading volume  \parencite{petukhina_rise_2021}. This lays groundwork for the calibration of crypto option pricing models as employed in \textcite{hou_pricing_2020}. By relating investor attention to the occurrence of jumps we confirm that it is indeed a predictor and show that this result applies even at the highest frequency. However, investigating the performance of available HF econometric tools reveals that the market microstructure of cryptos is so specific that new methodologies for investigating jumps and measuring microstructure noise in crypto markets are required. 

We add to previous high frequency jump analyses very recent data acquired from the \protect \includegraphics[height=0.5cm]{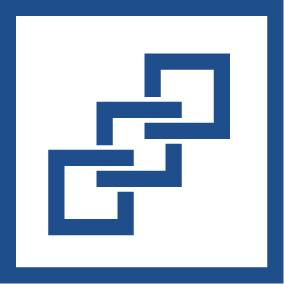}  {\color{blue}\href{https://blockchain-research-center.de}{Blockchain Research Center}} that captures the latest period of increased market activity of the largest CCs. We account for the interconnectedness of crypto markets by looking at some of the largest exchanges in North America, Europe and Asia. Our dataset consists of observations from April 12, 2019 until February 8, 2021. It consists of tick data collected from the exchanges Binance, Bitfinex, Bitstamp, Coinbase Pro, HitBTC, OKex and Poloniex. It features the currencies Bitcoin (BTC), Bitcoin Cash (BCH), Ethereum Classic (ETC), Ethereum (ETH), Litecoin (LTC) and Ripple (XRP). Using this new dataset we present the following empirical findings.

In total, we detected 1,046 jumps over all assets, of which ca. 68\% were negative. Even though negative jumps dominate in quantity, the overall distribution of jump sizes is still positively skewed. This is partly due to observed time frame that includes the effects of the recent Covid  crisis. Nevertheless, these results confirm findings on the number and distribution of jumps in earlier periods. While CCs tend to jump more than traditional assets, they still share some common properties: jumps seem to have a stronger impact during bull markets than during crashes. This becomes especially evident when linking jumps to important events. The results strongly suggests that jumps are clustered around periods of high investor attention.

The largest assets BTC and ETH jump on 66\% (58\%) of all observed trading days. Smaller currencies have less testing days as they are often too illiquid for a high frequency evaluation. This makes the interpretation of their test results difficult. In general, higher capitalized currencies tend to jump more, but more observations do not automatically mean more jumps. Next to the assumptions on high frequency data, the theoretical properties of the methodologies we use have been derived from traditional assets. Consequently, jump detection in CCs is less accurate due to differences in market microstructure noise. This calls for new approaches for noise robust volatility and jump estimators for CCs.

The jumps we detected show seasonal patterns. We detect more jumps in the middle of the week than on weekends, and most jumps during 1pm-5pm UTC with a sharp drop from 1am-6am UTC. This indicates that seasonality patterns in jumps are similar to seasonality patterns in volatility and trading volume. Therefore, a connection between these variables seems to exist.
Even though we observe several extreme jumps of more than +-10\% in a single moment, small jumps of less than +-2.5\% are dominating. They account for roughly 87\% of all jumps. We observe only 11 positive and 7 negative extreme jumps, making these events very rare. Note that we distinguish between extreme jumps that are detected using HF econometrics tools, and extreme returns that are unprocessed log returns, either daily or in HF. 

To investigate the relationship between jumps and returns we run regressions where we introduce a dummy variable for separating days with and without jumps, and additional dummy variables for days with positive (negative) jumps. We then regress end of day returns against these variables. We find a significant influence of these intra-day jumps on end-of-day returns. In case of a positive jump on a specific trading day, the end-of-day return is also likely to be positive and vice versa. In case of an extreme intra-day jump, it is likely that we will also observe an extreme return. Since CCs are traded 24/7, end-of-day returns are less meaningful. Therefore, we additionally regress intra-day jumps against end-of-day returns of the following trading day and find no evidence for an effect on the following day. In conclusion, HF jumps are an important driver of CC prices in the short run, and need to be accounted for in any meaningful option pricing model.

The remainder of this paper is structured as follows: Section \ref{sec:lit_review} discusses related literature and gives an overview of important events in the crypto universe. Section \ref{sec:methodology} describes the methodology, including a problem statement, which is jump detection in high frequency markets. It describes the basic model and the applied jump-testing procedures. Section \ref{sec:data} describes the data and its properties. Section \ref{sec:results} discusses the empirical results, like number of jumps, jump sizes, daily and weekly seasonality in jumps, and the effect of jumps on returns. Section \ref{sec:conclusion} concludes this paper.

\section{Literature Review} \label{sec:lit_review}
The widely unexplored properties of CCs have attracted the interest of researchers. CCs are distinct from other financial assets, commodities, or currencies. They can be traded 24/7, are largely unregulated, and currently highly speculative because the technology behind them is still in development. The promise of high returns in relatively short time frames, and the interest in their use case as a new, digital kind of currency has led to increased market capitalization. Bitcoin alone has a market capitalization of more than \$1,100 bn. (October 15, 2021), and has grown bigger than high profile stocks in the S\&P500 like Tesla and Facebook, only topped by the four largest tech stocks Apple, Amazon, Microsoft and Alphabet. 

This research builds upon previous work on jumps in CCs. The lack of fundamental research on their properties along with the availability of trading data from major exchanges on a tick level make CCs an interesting object of study, hence we focus on them in this paper as an example for digital assets in general. We add jumps as an additional explanatory variable on returns to \textcite{liu_risks_2020}. 
A large branch of literature on the definition and detection of jumps exists. \textcite{christensen_fact_2014} show a summary of recent approaches for detecting them. They find that, historically, jumps have been detected mainly in frequencies as low as daily. Even high frequency analyses mostly looked at intervals like 5 minutes, not at tick data. Possibly, this is because many assets are not assumed to be traded frequently enough to justify a HF analysis. \textcite{mukherjee_chapter_2020} provide an overview for jump tests and also co-jump tests. In this paper, we employ the methodology of \textcite{lee_jumps_2012}, since it allows for moment based detection of jumps and is robust to market microstructure noise. To lower the probability of spurious jump detection, we additionally test for jumps using the approach as proposed in \textcite{ait-sahalia_testing_2012} and only accept jumps detected by Lee \& Mykland on days where Ait-Sahalia, Jacod and Li also detected a jump. We address the issue of multiple testing by using a Bonferroni correction on both tests.

Researchers have analyzed CCs, e.g. by looking at correlations between sentiment and prices \parencite{aste_cryptocurrency_2019} and correlations between different CCs \parencite{hardle_understanding_2020}, studying them from a monetary perspective \parencite{yermack_chapter_2015}, or putting them into indices such as CRIX ({\color{blue}\href{https://thecrix.de/}{thecrix.de}}) \parencite{trimborn_crix_2018}. \textcite{burnie_analysing_2020} finds that social media discussions can be the trigger for price shifts in CCs. \textcite{makarov_trading_2020} find that cross-country arbitrage opportunities frequently appear in CCs. \textcite{menkveld_flash_2018} study the connection of Flash-Crashes and cross-arbitraging in high frequency, but on the example of traditional assets, whereas \textcite{borri_conditional_2019} outlines tail-risks of CCs. \textcite{trimborn_investing_2020} show that the risk-return trade-off can be improved by diversifying the digital asset portfolio. \textcite{giudici_cryptocurrencies_2020} discusses recent trends in academic discussions on digital assets. Certainly, a recent trend is to use available indices for building option pricing models \parencite{madan_advanced_2019}. Analyses on the performance of indices such as in \textcite{chen_first_2016, elendner_chapter_2018} help researchers calibrate their models and portfolio composition. Given the constant emergence of new assets, \textcite{howell_initial_2020} explores success factors for new launches.

Studying financial assets in high frequency (HF) poses new challenges. \textcite{ait-sahalia_high-frequency_2014} outline that high frequency data has unique properties due to irregularly spaced observations resulting in asynchronity, market microstructure noise, and information loss if data is aggregated to one-second-intervals. Heavy tails, long memory in volatility, intra-day and intra-week seasonality, as in other financial assets additionally result in non normality. The concept of market microstructure noise as introduced e.g. by \textcite{black_noise_1986}, and with respect to high frequency and algorithm trading as in \textcite{ohara_market_1998} calls for robust volatility and jump devices. Research on jumps also suggested that correlated assets often jump together, see \textcite{ait-sahalia_modeling_2015}, and \textcite{caporin_systemic_2017}. In addition, the jump sizes may be correlated in assets and in volatility, as indicated by \textcite{hou_pricing_2020}.

\begin{figure}[ht]
\centering
\begin{subfigure}{.3\textwidth}
\includegraphics[width=\textwidth]{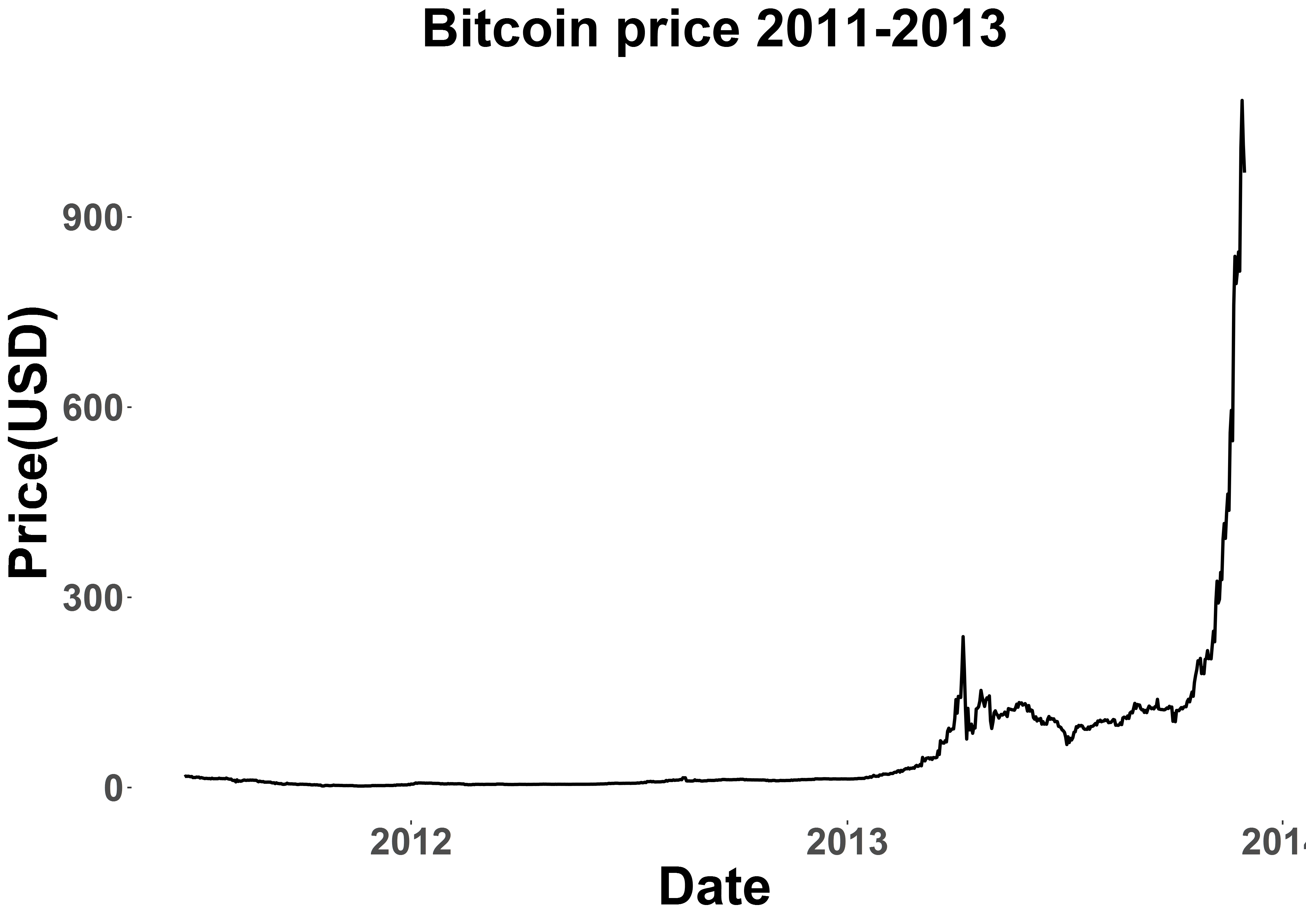} 
\end{subfigure}
\begin{subfigure}{.3\textwidth}
\includegraphics[width=\textwidth]{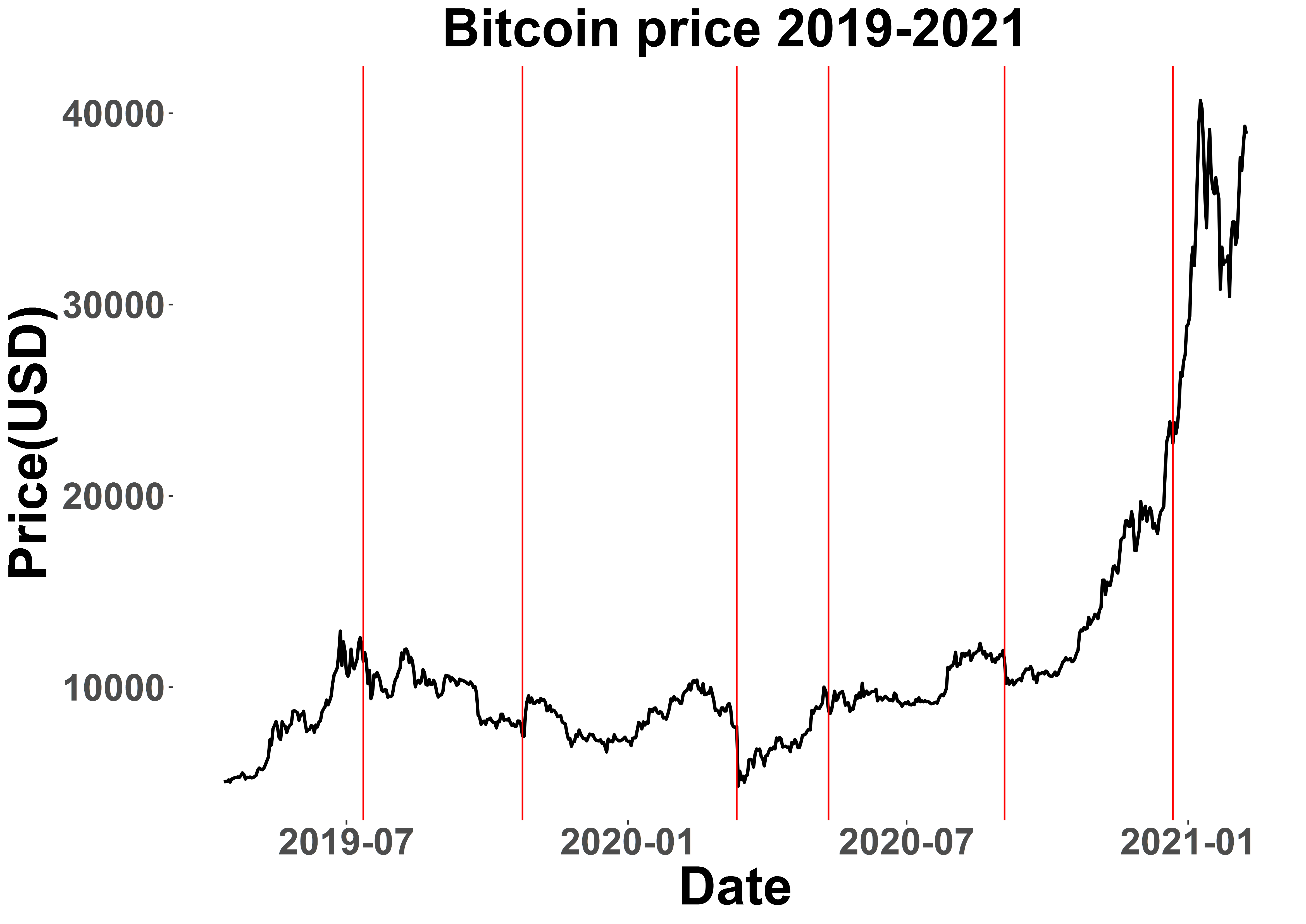}
\end{subfigure}

\begin{subfigure}{.3\textwidth}
\includegraphics[width=\textwidth]{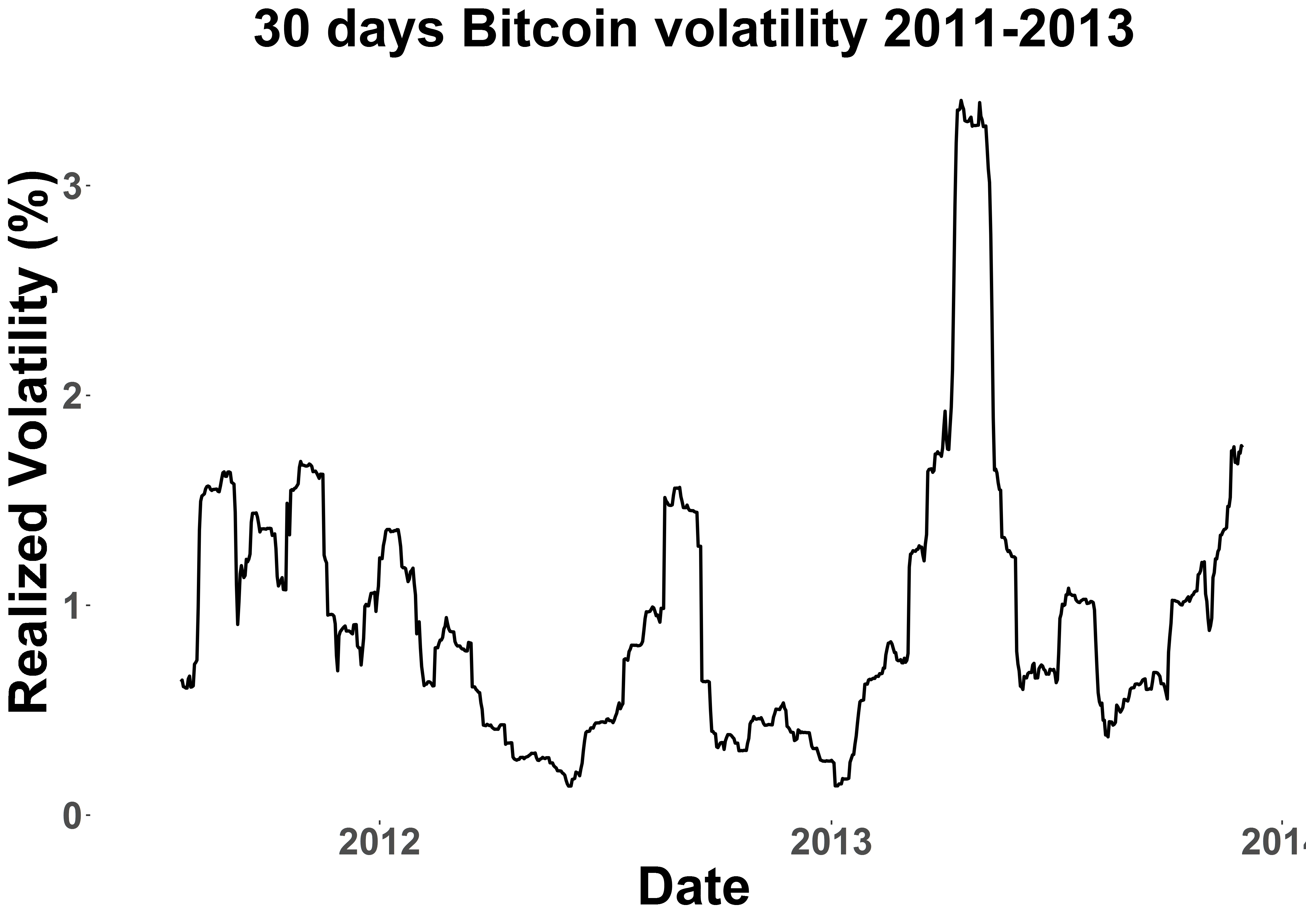}
\end{subfigure}
\begin{subfigure}{.3\textwidth}
\includegraphics[width=\textwidth]{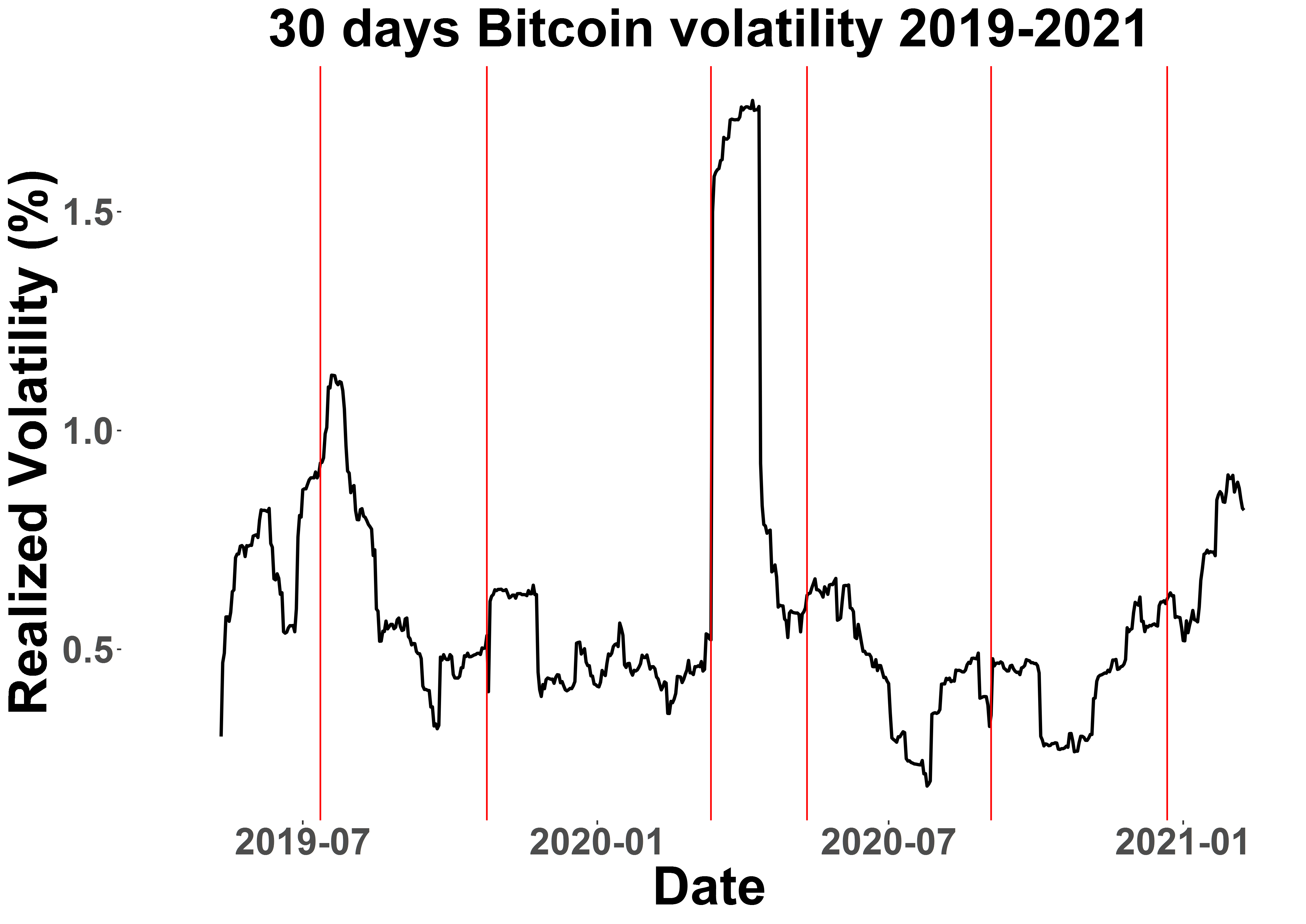}
\end{subfigure}

\begin{subfigure}{.3\textwidth}
\includegraphics[width=\textwidth]{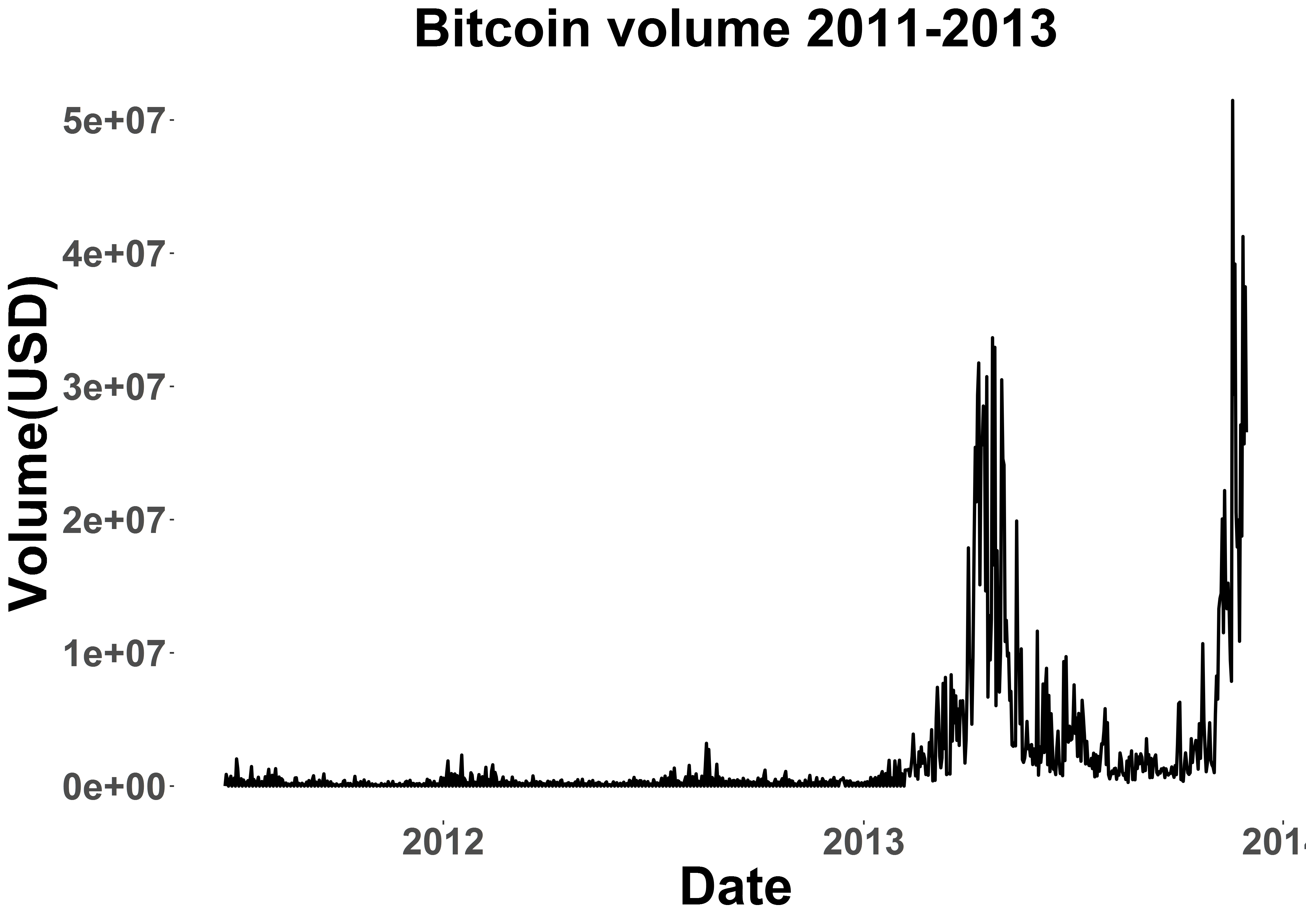}
\end{subfigure}
\begin{subfigure}{.3\textwidth}
\includegraphics[width=\textwidth]{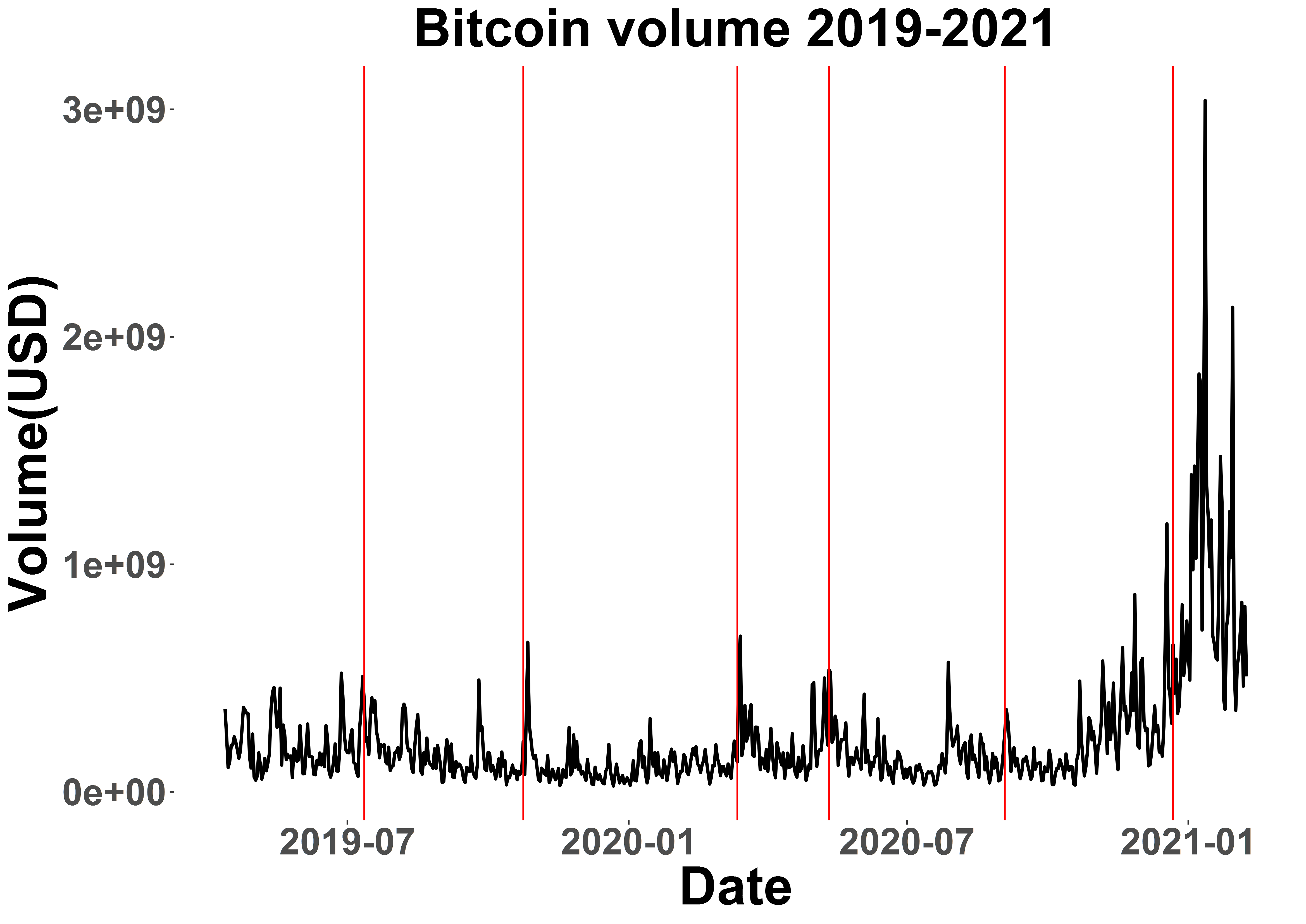}
\end{subfigure}

\caption{Bitcoin price, realized volatility and volume. (Source: Quandl) \protect \includegraphics[height=0.5cm]{plots/qletlogo_tr.png} {\color{blue}\href{https://github.com/QuantLet/JumpDetectR}{JumpDetectR}}
}
\label{fig:btctimechange}
\end{figure}

Adding to previous literature on jumps in CCs, e.g. \textcite{scaillet_high-frequency_2020} who employs a jump detection technique on Bitcoin prices for the period of 2011-2013, we follow a similar approach for the most recent period that includes a temporary market crash due to the Covid-Crisis and a bull run to new all-time-highs for Bitcoin. Figure \ref{fig:btctimechange} shows the dynamics of Bitcoin prices, trading volume, and realized volatility both in 2011-2013 and 2019-2021. Price and volume have risen exponentially, while realized volatility has decreased. The red lines indicate a selection of important events in the crypto universe during our observation period: 
\begin{enumerate}
    \item Former US President Donald Trump releases a series of Tweets against BTC and other cryptos ({\color{blue}\href{https://corpgov.law.harvard.edu/2019/08/06/a-roadmap-for-president-trumps-crypto-crackdown}{12:15am UTC, July 12, 2019}}),
    \item Chinese President Xi Jinping announces his support for blockchain technology ({\color{blue}\href{http://www.gov.cn/xinwen/2019-10/25/content_5444957.htm}{10:13am UTC, October 25, 2019}}),
    \item In the course of the Covid crisis, US stock markets experienced so called Black Thursday, causing large sell-offs and trading got suspended for 15 minutes on NYSE. This affected the crypto market as well. ({\color{blue}\href{https://www.cnbc.com/2020/03/12/stock-futures-hit-a-limit-down-trading-halt-for-a-second-time-this-week-heres-what-that-means.html}{1:35pm UTC, March 12, 2020}}),
    \item The third BTC halving comes into effect, which halves the reward for mining a BTC block. Historically, this event has led to major bull runs in the nearer future ({\color{blue}\href{https://www.blockchain.com/btc/block/000000000000000000024bead8df69990852c202db0e0097c1a12ea637d7e96d}{7:30pm UTC, May 11, 2020}}),
    \item Notable movements to large exchanges could be observed, indicating that some whales sold significant amounts of Bitcoin at once ({\color{blue}\href{https://www.blockchain.com/btc/tx/f0712a1b1faa5d6a8f45767dced176f7fb8d2e58b4e20355007c95ef954cd15f}{1:49pm UTC, September 3, 2020}}),
    \item The SEC files a lawsuit against Ripple, arguing that XRP is a security. Consequently, major exchanges suspend trading activities for XRP ({\color{blue}\href{https://web.archive.org/web/20201227204540/https://www.sec.gov/news/pressreleases.rss}{10:13pm UTC, December 22, 2020}}).
\end{enumerate}

We investigate how these events relate to the occurrence of jumps. Clearly, in terms of price, realized volatility and trading volume these events had a significant influence on BTC and thus indicate a connectedness of jumps to important events, and with volatility and trading volume. 

\section{Methodology} \label{sec:methodology}

\subsection{Basic idea} \label{sec:problemstatement}
Jumps are omnipresent in financial time series, but not well explored when it comes to CC time series, especially not in high frequency. For investors it is important to get reliable expectations on the frequency of jumps and their size in order to minimize the possibility of losses from large downside risks and ideally participate in profitable bull runs. To extract jumps from HF CC time series, the methodology of \textcite{lee_jumps_2012} is employed. We keep only those jumps for which the methodology of \textcite{ait-sahalia_testing_2012} also detected a jump on the same day. These methods make use of the pre-averaging approach to denoise the time series, as well as jump and noise robust volatility estimates for the usual variation in every sample. Econometric methods for determining the right sampling frequency were introduced and discussed e.g. in \textcite{ait-sahalia_how_2005, ait-sahalia_ultra_2011, zhang_tale_2005, liu_does_2015, jacod_statistical_2017, li_unified_2018}. Important contributions to pre-averaging methods are \textcite{jacod_microstructure_2009}, \textcite{jacod_limit_2010} and \textcite{christensen_pre-averaging_2010}. The theory was further developed by \textcite{hautsch_preaveraging-based_2013}, \textcite{podolskij_edgeworth_2017}, and \textcite{li_dependent_2020} among others. 
To estimate the variation of HF data, \textcite{barndorff-nielsen_power_2004, barndorff-nielsen_econometrics_2006} and \textcite{barndorff-nielsen_limit_2006} proposed taking multipower variation. Robustifications can be found e.g. in \textcite{vetter_limit_2010}. The advances in HF econometrics allow us thus to employ various techniques for handling the difficulties of our underlying data and to test for jumps in CCs.

We test each currency on each exchange for every day separately using standard tools of HF econometrics. To address the issue of multiple testing we use a Bonferroni correction. Recall figure \ref{fig:btctimechange}. From these HF time series we want to determine all jumps to answer these questions:

\begin{enumerate}
    \item How often do jumps occur? Can they be related to important events?
    \item Which CCs jump the most and what distinguishes them?
    \item How accurate are our jump detection methods compared to traditional stocks?
    \item When do jumps occur? Are there any intra-day, weekly seasonality patterns?
    \item How are jump sizes distributed? Do big or small jumps dominate?
    \item Do intra-day jumps have a significant effect on end of day returns? Is there a difference between positive and negative jumps?
\end{enumerate}
It is commonly known that relevant events such as halving or Tweets by influential people can cause increased volatility and trading volume. By answering these questions, we show that HF jumps significantly drive the price process of cryptos, have similar seasonality patterns as volatility and volume and are clustered in time around notable incidents. Thus we are laying out important characteristics for crypto option pricing models that include a well-specified jump component.

\subsection{Modeling jumps} \label{sec:model}
Consider a complete probability space $(\Omega, \mathcal{F}_t, \mathbb{P})$ with $\Omega$ the set of all possible events in the CC market, ${\mathcal{F}_t:t\in[0,T]}$ a right-continuous information filtration and $\mathbb{P}$ the probability measure. Now define the model 
\begin{equation}
    dX_{t} = \sigma dW_{t} + Z_{t}dJ_t,
\end{equation}
where $t\in [0,T]$, an arbitrary point in time within one trading day.
Note that since we observe CCs, a trading day has 24 hours. $X_t\subset \mathbb {R}$ is the log price at all times, $\sigma\in \mathbb {R}^+ $ denotes a volatility estimate assumed to be constant over the observed period. $W_t\subset \mathbb {R}$ is a Brownian motion, $J_t\in \left \{ 0,1 \right \}$ denotes the jump arrival indicator with jumps of size $Z_t \subset \mathbb {R}$. Our objective is to find $J_t$ in a data driven way. This is a fairly simple model. Extensions like the SVCJ will be discussed throughout.

\subsection{Lee \& Mykland jump test} \label{sec:lmtest}
To identify jumps in HF data, a common approach is to use a pre-averaging method to reduce noise and to calculate some multipower variation to get an estimate of governing volatility in the observed process. This method allows for moment based detection of intra-day jumps. The observed, with market microstructure noise $\epsilon$ contamined price at time $t$ is
\begin{equation*}
\tilde{P_{t}} = X_{t} + \epsilon_t.
\end{equation*}

Note that $\epsilon_t$ has a mean of $0$ and a variance of $q^2$ where $q$ is the market quality parameter that describes the degree of market imperfection. Let $k$ be obtained empirically via the empirical autocorrelation function lag order of the observed price, and $M$ the block size which we choose as per recommendation of \textcite{lee_jumps_2012}.  We compute the average price over the block size $M$
\begin{equation*}
\hat{P}(t_j) \overset{\text{def}}{=}  M^{-1}\sum_{i = \left \lfloor j/k\right \rfloor}^{\left \lfloor j/k\right \rfloor + M-1}\tilde{P}(t_{ik}),
\end{equation*} 
with $j = kM, 2kM, 3kM, ...$ and calculate the asymptotically normal test statistic 
\begin{equation*}
\bar{P}(t_j) \overset{\text{def}}{=}  {\hat{P}(t_{j+kM})- \hat{P}(t_{j})}
\end{equation*}
to determine whether a jump has happened between $t_{j}$ and $t_{j+kM}$. The test statistic converges in distribution to \begin{equation*}
B_n^{-1} \Bigl( \frac{\sqrt{M}}{\sqrt{V_n}} | \bar{P}(t_j) | - A_n \Bigr) \overset{\mathcal{L}}{\rightarrow}\xi,
\end{equation*} 
where $\xi$ follows a standard Gumbel distribution, $V_n \overset{\text{def}}{=}  {\mbox{Var}\left [ \sqrt{M}\bar{P}(t_j)\right ]}$ is the scaled variance of the test statistic and
\begin{equation*}
A_n = \left ( 2 \log\left \lfloor \frac{n}{kM} \right \rfloor \right )^{1/2} - \frac{\log\pi+\log\left ( \log\left \lfloor \frac{n}{kM} \right \rfloor \right )}{2\left (2\log\left \lfloor \frac{n}{kM} \right \rfloor  \right )^{1/2}}, 
\end{equation*}
\begin{equation*}
B_n = \frac{1}{\left (2\log\left \lfloor \frac{n}{kM} \right \rfloor  \right )^{1/2}},
\end{equation*}
under the null hypothesis of no jumps we can then say that if e.g. $\hat \xi >$ 99th percentile of the standard Gumbel distribution we observe a jump. A more formal description of the method can be found in the appendix in section \ref{sec:app_lmjumptest}.

\subsection{Aït-Sahalia, Jacod and Li jump test} \label{sec:ajltest}
This method again uses pre-averaging to obtain a noise robust power variation $\bar{V}$ for any observed log return time series $Z$. The ratio of two differently weighted power variation converges to some finite limit under the null hypothesis of no jumps.

To compute the robustified test statistic for jumps we set up the constants

\begin{equation*}
    \gamma = \frac{(\bar g)(2)}{(\bar h)(2)},  \gamma^{'} = \frac{(\bar g)(p)}{(\bar h)(p)},  \gamma^{''} = \frac{\gamma^{p/2}}{\gamma^{'}},
\end{equation*}
under the assumption that $\gamma^{''} >1$, where $g$ and $h$ are pre-averaging weights, and $p=4$. The noise robust test statistic then
\begin{equation}
    S_{RJ}(g,h,p)_n = \frac{\bar{V}(Z,g,p)_T^n}{\bar{V}(Z,h,p)_T^n},\label{eq:B_xpd}
\end{equation}
with $n$ the number of observations, and $T$ the time horizon, in our case a full day of observations.
Asymptotically, the test statistic has the following limit behavior:
\begin{equation}
S_{RJ} \left ( p,k,\Delta_n \right ) \overset{\mathbb{P}}\rightarrow \left\{\begin{matrix}
1 & \text{on } \Omega_T^j  \\ 
\gamma^{''} & \text{on } \Omega_T^c  
\end{matrix}\right.
\end{equation}
and the critical value for rejecting the null hypothesis of no jumps is 
\begin{equation}
C_n^c =  \Bigl\{S_{RJ}(g,h,p)_n <  \gamma^{''}- z_{\alpha} \Delta_n^{1/4} \sqrt{\Sigma_{RJ,n}^c} \Bigr\},
\end{equation}
where $\Sigma_{RJ,n}^c$ is the scaled variance of the test statistic, $\Delta_n$ denotes the $n$th difference of observations, and $ z_{\alpha}$ is the corresponding quantile of the standard normal distribution, s.t. the rejection region can be obtained by choosing a respective $\alpha$. Further details and a more formal explanation of the method can be found in the appendix in section \ref{sec:app_ajljumptest}.
\FloatBarrier
\section{Data} \label{sec:data}
The dataset is collected for the time frame April 12, 2019 until February 8, 2021.  It contains a total of 1,039,407,036 observations of ticks aggregated to frequencies of 1, 5, 10, or 15 seconds for testing, depending on whether or not the observed interval contains at least $95\%$ of the observations at these frequencies.  A dataset aggregated to 1 second intervals would have to have at least $0.95*86,400=82,080$ observations. If the dataset had a frequency of less than of 15 seconds, we concluded that the respective data is not "high frequency“ and therefore omitted the data. We did not calculate a test statistic on the respective days in order to avoid violating the assumptions of HF data in the testing methodologies. We admit that this threshold seems arbitrary, but needed to decide on a cutoff point as there is no unique definition of what exactly is a high frequency dataset. We sampled price data of six of the largest currencies in terms of market capitalization from seven exchanges in Europe, Asia and US. To account for cross-market arbitraging, we aggregate the prices per symbol over the different exchanges and take the mean price for each point in time in case of simultaneous observations. As usual in HF literature, we remove bounceback outliers and returns outside of a range of 10 standard deviations. In terms of the amount of standard deviations the literature varies. \textcite{lee_jumps_2012} use for example a cutoff of 7 standard deviations. To account for the high volatility of CC we relax this cutoff slightly. The data is obtained from the \protect \includegraphics[height=0.5cm]{plots/brc.png}  {\color{blue}\href{https://blockchain-research-center.de}{Blockchain Research Center}}.

\begin{figure}[ht]
    \centering
    \includegraphics[width=0.75\textwidth]{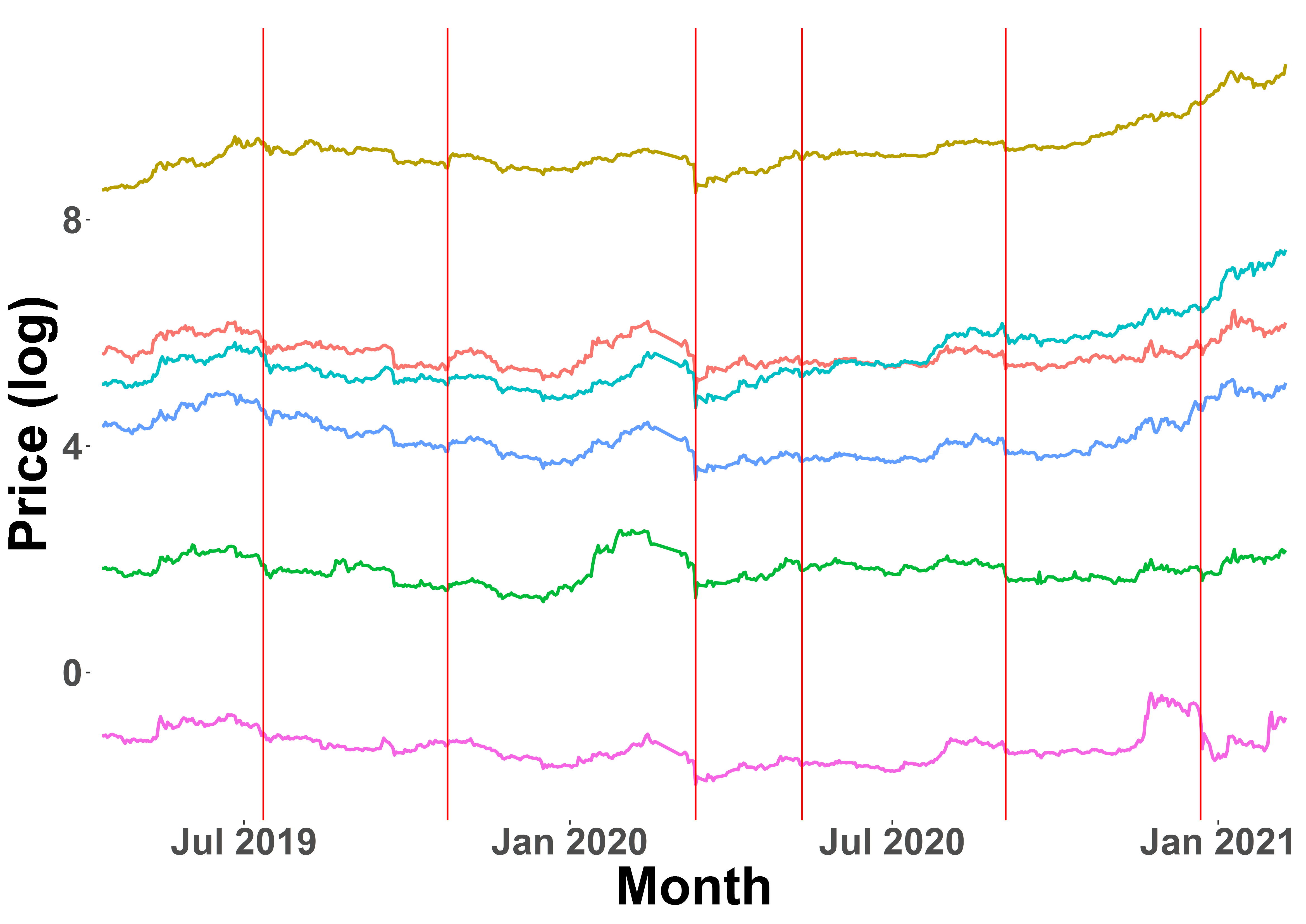}
    \caption{Log price of {\color[HTML]{B79F00} BTC}, {\color[HTML]{00BFC4} ETH}, {\color[HTML]{F8766D} BCH}, {\color[HTML]{619CFF} LTC}, {\color[HTML]{00BA38} ETC}, and {\color[HTML]{F564E3} XRP}. \protect \includegraphics[height=0.5cm]{plots/qletlogo_tr.png} {\color{blue}\href{https://github.com/QuantLet/JumpDetectR}{JumpDetectR}}}
    \label{fig:daily_p_all}
\end{figure}

Figure \ref{fig:daily_p_all} shows the aggregated daily prices of the six CCs on a log scale along with a selection of relevant events in the crypto universe. Recall that these are a series of Donald Trump Tweets attacking cryptos in July 2019, a pro-blockchain statement of Chinese President Xi Jinping in October 2019, the peak of the Covid crisis in March 2020, the third BTC halving in May 2020), large scale wallet movements to exchanges in September 2020), and a lawsuit against Ripple in December 22, 2020. The plot indicates that these events have an influence not only on BTC, but also on other cryptos in form of sudden price changes and increased volatility. To understand the nature of these price changes better, we will now give an overview of the data set and then turn to the summary statistics of daily and HF returns and to the frequency of extreme returns.

\FloatBarrier

Table \ref{n_per_id} shows the number of observations per exchange.  With 575 million observations, 55.3\% of observations were collected from Binance. OKex and Coinbase Pro follow with large distance. In contrast, only 15 million (1.5\%) observations were collected from the smallest exchange Poloniex.
Since we collected all ticks, the number of observations resembles the frequency of trades happening on each market place. These differences have various reasons. E.g., not all symbols are traded on all exchanges, and some exchanges are more popular in certain countries. For example, Coinbase Pro and Bitstamp allow for trading CCs against EUR, thus making them more popular in Europe. As a consequence, we observe a high level of concentration on only few exchanges. 

\begin{table}[ht] 
\begin{subtable}{\linewidth}
\centering
\begin{tabular}{@{\extracolsep{5pt}} lr} 
\\[-1.8ex]\hline 
\hline \\[-1.8ex] 
Exchange & N \\ 
\hline \\[-1.8ex] 
Poloniex & $15,579,586$ \\ 
Bitstamp & $33,350,101$ \\ 
Hitbtc & $39,371,798$ \\ 
Bitfinex & $81,544,092$ \\ 
Coinbase Pro & $112,063,403$ \\ 
OKex & $182,532,074$ \\ 
Binance & $574,965,982$ \\ 
\\[-1.8ex]\hline 
\hline \\[-1.8ex] 
\end{tabular} 
  \caption{N obs. per exchange.} 
  \label{n_per_id} 
\end{subtable}

\begin{subtable}{\linewidth}
\centering
\begin{tabular}{lrrrrr}
\\[-1.8ex]\hline 
\hline \\[-1.8ex] 
Symbol & N (raw) &  \% N (raw) & N (aggregated) & \% N (agg.) & \% N vs raw \\ 
  \hline
ETC & 27,875,817 & 2.68 & 3,111,940 & 3.12 & 11.16 \\ 
  BCH & 37,803,009 & 3.64 & 3,938,423 & 3.94 & 10.42 \\ 
  LTC & 55,142,174 & 5.31 & 6,258,789 & 6.27 & 11.35 \\ 
  XRP & 88,228,936 & 8.49 & 15,846,840 & 15.87 & 17.96 \\ 
  ETH & 159,238,124 & 15.32 & 26,232,067 & 26.27 & 16.47 \\ 
  BTC & 671,118,976 & 64.57 & 44,469,618 & 44.53 & 6.63 \\ 
\\[-1.8ex]\hline 
\hline \\[-1.8ex] 
\end{tabular}
\caption{ \label{n_obs_s} N obs. per symbol.} 
\end{subtable}
\caption{Data overview.} 
\end{table} 

Table \ref{n_obs_s} shows the number of observations per symbol in the raw dataset and after aggregation. Column two shows the total number of observations in the raw dataset. Column three shows the percentage of observations that each currency contributes to the total number of observations in the dataset. We observe that, similarly to the number of observations per exchange, the data is highly concentrated on a few currencies. In terms of currencies, the market is thus even stronger concentrated. Column four and five show the number and the percentage of observations after aggregation. Column six shows the percentage of observations that are left after aggregation. Since we aggregate prices over different exchanges and only on days with enough data, the amount of observations per symbol is reduced drastically. The data pool has then a total of 99,857,677 million observations on a total of 645 trading days. This is slightly lower than the full amount (669 days) caused by database outages or dropped data due to quality issues. After aggregation, the concentration reduces substantially, as we can have no more than 86,400 observations on a single trading day on the highest possible frequency of one second. Therefore, BTC is less dominant in the aggregated data, and while ETH and XRP gain in shares of observations, the smaller three currencies barely change in percentage share of total observations. We observe that no currency keeps more than 18\% of observations (XRP), while BTC only keeps slightly more than 6.5 \% of observations after aggregation. 

Table \ref{ret_secondly} shows the returns of aggregated CC prices on the highest frequency. As often in financial data, the observed log returns are not normally distributed, as the higher moments show in the last two columns. The large kurtosis shows that there are many values close to zero. This is likely due to the presence of market microstructure noise in HF data, which can be eliminated e.g. by sampling at lower frequencies or in many cases by means of using techniques such as pre-averaging. The summary statistics indicate that even in the highest possible frequency large positive or negative returns of up to +-37\% (BCH) and +-34\% (XRP) are observed. The minima and maxima seem symmetric. This is only partly due to the cutoff in the range of 10 standard deviations. In fact, only ETC and LTC are affected by this cutoff where the minimum for ETC (LTC) is at -101\% (-83\%), and the maximum for ETC (LTC) is at +78\% (+82\%) for a single return. Unlike all other CCs with a positive skewness, BTC shows a negative skewness. The presence of market microstructure noise makes the interpretation of these returns even more difficult. However, as table \ref{extreme_returns} shows, extreme returns are not outliers, but commonly observed even in HF. In total, we observe more than 4,800 returns smaller than -5\% and more than 4,800 returns larger than +5\%. 229 observed HF returns are smaller than -10\% and 231 returns are larger than +10\%. Looking at returns of +- 20\%, we still observe 34 (34) returns, whereas 17 (16) returns are +-30\%. Surely, these returns cannot be explained by market microstructure noise and indicate that extreme returns have a significant influence on the price process. To see how these extreme returns fit into the big picture, we look at daily returns next.

\begin{table}[ht]
  \begin{subtable}{\linewidth}
\centering
\begin{tabular}{lrrrrrrrr}
\\[-1.8ex]\hline 
\hline \\[-1.8ex] 
Currency & Min. & 1st Qu. & Median & Mean & 3rd Qu. & Max. & Skewness & Kurtosis \\ 
  \hline
BCH & -0.37 & -0.0003 & 0 & 0 & 0.0003 & 0.37 & 0.06 & 1,159.73 \\ 
  BTC & -0.20 & -0.0001 & 0 & 0 & 0.0001 & 0.20 & -0.04 & 768.36 \\ 
  ETC & -0.10 & -0.0003 & 0 & 0 & 0.0003 & 0.10 & 0.06 & 1,028.99 \\ 
  ETH & -0.25 & -0.0002 & 0 & 0 & 0.0002 & 0.24 & 0.00 & 1,003.80 \\ 
  LTC & -0.19 & -0.0003 & 0 & 0 & 0.0003 & 0.19 & 0.11 & 407.80 \\ 
  XRP & -0.34 & -0.0002 & 0 & 0 & 0.0003 & 0.34 & 0.46 & 2,183.87 \\ 
\\[-1.8ex]\hline 
\hline \\[-1.8ex] 
\end{tabular}
\caption{  \label{ret_secondly} Returns per symbol: highest frequency.} 
\end{subtable}

\begin{subtable}{\linewidth}
\centering
\begin{tabular}{lrlr} 
\\[-1.8ex]\hline 
\hline \\[-1.8ex] 
Negative & Counts & Positive & Counts  \\ 
\hline \\
$<$ -0.05 & 4,804 & $>$ 0.05 & 4,801 \\
$<$ -0.1 & 229 & $>$ 0.1 & 231 \\ 
$<$ -0.2 & 34 & $>$ 0.2 & 34  \\ 
$<$ -0.3 & 16 & $>$ 0.3 & 17  \\ 
\\[-1.8ex]\hline 
\hline \\[-1.8ex] 
\end{tabular} 
\caption{\label{extreme_returns} Extreme returns: highest frequency.} 
\end{subtable}
\caption{Return statistics in high frequency.} 
\end{table}

\begin{figure}[ht]
    \centering
    \includegraphics[width=0.75\textwidth]{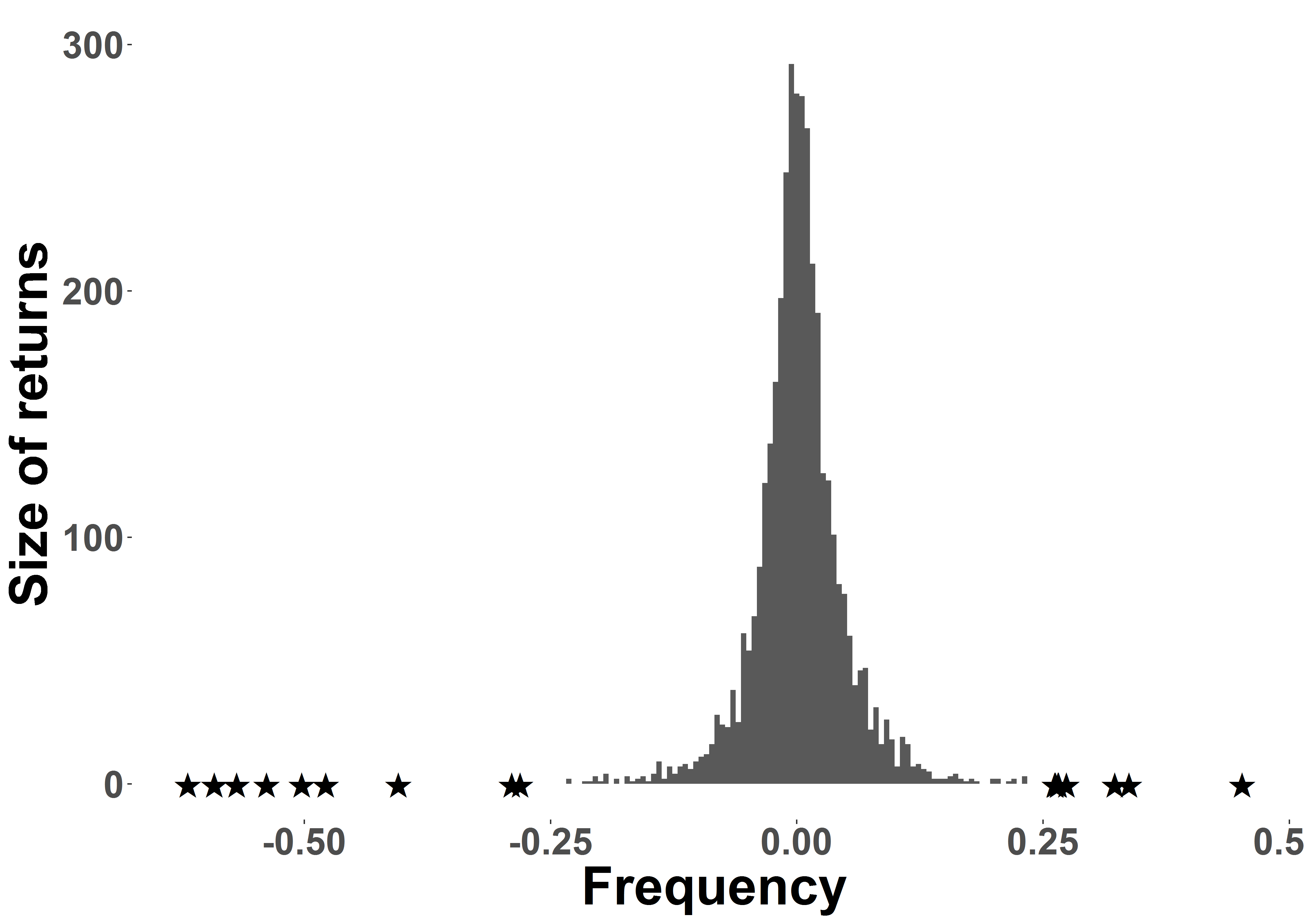}
    \caption{Histogram of daily returns on all time series. \protect \includegraphics[height=0.5cm]{plots/qletlogo_tr.png} {\color{blue}\href{https://github.com/QuantLet/JumpDetectR}{JumpDetectR}}}
    \label{fig:ret_daily}
\end{figure}

Table \ref{ret_daily} shows the statistics of the daily returns of CCs and figure \ref{fig:ret_daily} shows their histogram. In contrast to HF returns, we observe a negative skewness in all time series (this could however change already given the most recent bull market since February 2021). The kurtosis values are much lower, as the daily frequency eliminates market microstructure noise. Furthermore, the minima and maxima have changed significantly. While the minima become lower for all currencies, the maxima become more extreme for some CCs, and less extreme for others. Thus, the extreme returns on a daily scale in comparison to HF returns seem to differ in some cases, but seem highly similar in others. Intuitively, such extreme events in HF should also have an influence on the price process in the longer run given their disruptive nature (they would be highly unlikely in a pure random walk model). These findings lead us to investigate whether extreme daily returns could be explained by intra-day singularities. In such a scenario, only few large observations would cause extreme daily returns and we seek to identify these returns by means of a jump testing procedure. The histogram shows that while these extreme minima are tail events that happened during the covid shock, there is a substantial amount of returns that are larger than +-5\%.

Table \ref{extreme_returns_daily} shows the number of extreme returns on a daily basis. Naturally, we observe less extreme returns, but they are still frequently observed, which indicates that HF events have an influence on the price in the longer run.

\begin{table}[ht]
  \begin{subtable}{\linewidth}
  \centering
\begin{tabular}{lrrrrrrrr}
\\[-1.8ex]\hline 
\hline \\[-1.8ex] 
Currency & Min. & 1st Qu. & Median & Mean & 3rd Qu. & Max. & Skewness & Kurtosis \\ 
  \hline
BCH & -0.62 & -0.02 & 0 & 0 & 0.02 & 0.27 & -1.83 & 26.68 \\ 
  BTC & -0.50 & -0.01 & 0 & 0 & 0.02 & 0.18 & -2.50 & 36.78 \\ 
  ETC & -0.57 & -0.02 & 0 & 0 & 0.02 & 0.27 & -1.82 & 24.88 \\ 
  ETH & -0.59 & -0.02 & 0 & 0 & 0.03 & 0.23 & -2.34 & 30.83 \\ 
  LTC & -0.48 & -0.02 & 0 & 0 & 0.02 & 0.23 & -1.13 & 15.72 \\ 
  XRP & -0.54 & -0.02 & 0 & 0 & 0.02 & 0.45 & -0.30 & 26.53 \\ 
\\[-1.8ex]\hline 
\hline \\[-1.8ex] 
\end{tabular}
\caption{\label{ret_daily} Returns per symbol: daily frequency.} 
\end{subtable}

\begin{subtable}{\linewidth}
\centering
\begin{tabular}{lclc} 
\\[-1.8ex]\hline 
\hline \\[-1.8ex] 
Negative & Counts & Positive & Counts  \\ 
\hline \\
$<$ -0.05 & 344 & $>$ 0.05 & 430 \\
$<$ -0.1 & 95 & $>$ 0.1 & 105 \\ 
$<$ -0.2 & 19 & $>$ 0.2 & 16  \\ 
$<$ -0.3 & 7 & $>$ 0.3 & 3  \\ 
\\[-1.8ex]\hline 
\hline \\[-1.8ex] 
\end{tabular} 
\caption{\label{extreme_returns_daily} Extreme returns: daily frequency.} 
\end{subtable}
\caption{Return statistics on a daily frequency.} 
\end{table}

The data shows that the characteristics of the different CCs are quite similar, despite their differences in technology and investor attention. The markets are highly concentrated, s.t. most observations are collected from big coins such as BTC and ETH and on big exchanges such as Binance, OKex, and Coinbase Pro. Both in daily and in high frequency, extreme returns occur frequently. Whereas the unprocessed HF returns are difficult to interpret due to market microstructure noise, the daily returns are negatively skewed with long tails. Note that the results on returns differ from e.g. \textcite{liu_risks_2020} because we observe a much shorter time span in a later point in time that includes the effects of the Covid crisis in March 2020 and the latest bull run in early 2021. Indeed, for developing profitable trading strategies and reliable models, we need to account for these statistical properties. Given the high magnitude of returns even at tick level, we have to focus on HF events to preserve the information that this data contains, as we already see that events in HF and in daily frequency seem to be connected. If we want to model these dynamics we need to incorporate knowledge about jumps. In the following section we will show that daily dynamics are largely driven by the occurrence of jumps in high frequency.
\FloatBarrier

\section{Understanding CC jumps} \label{sec:results}
To investigate the occurrence of jumps in HF, some preprocessing is necessary. Since the two testing methodologies differ, we preprocess the dataset differently for both methodologies. Lee \& Mykland works best on processing ticks as it does not make any assumptions on the time between two observations. We determine consecutive jump detections by tagging jumps that were detected within 10 moments after initial detection and remove them from the data.
Aït-Sahalia, Jacod and Li on the other hand works best on equispaced data. Hence, we sample data either at frequencies of 1,5,10, or 15 seconds. To maintain a regular structure, we impute missing observations by inserting the last observed price. We set $\alpha = 0.999$ and apply a Bonferroni correction to minimize the detection of spurious jumps. In total, we observed 1,046 jumps on all assets, where Lee \& Mykland detected a jump and Aït-Sahalia, Jacod and Li detected the presence of jumps on the same day. Note that the method is robust to varying $\alpha$. With $\alpha = 0.99$, we detect 1,357 jumps, setting $\alpha = 0.95$ we detect 1,644 jumps, and with $\alpha = 0.9$ 1,787 jumps.

\begin{figure}[ht]
    \centering
    \includegraphics[width=0.75\textwidth]{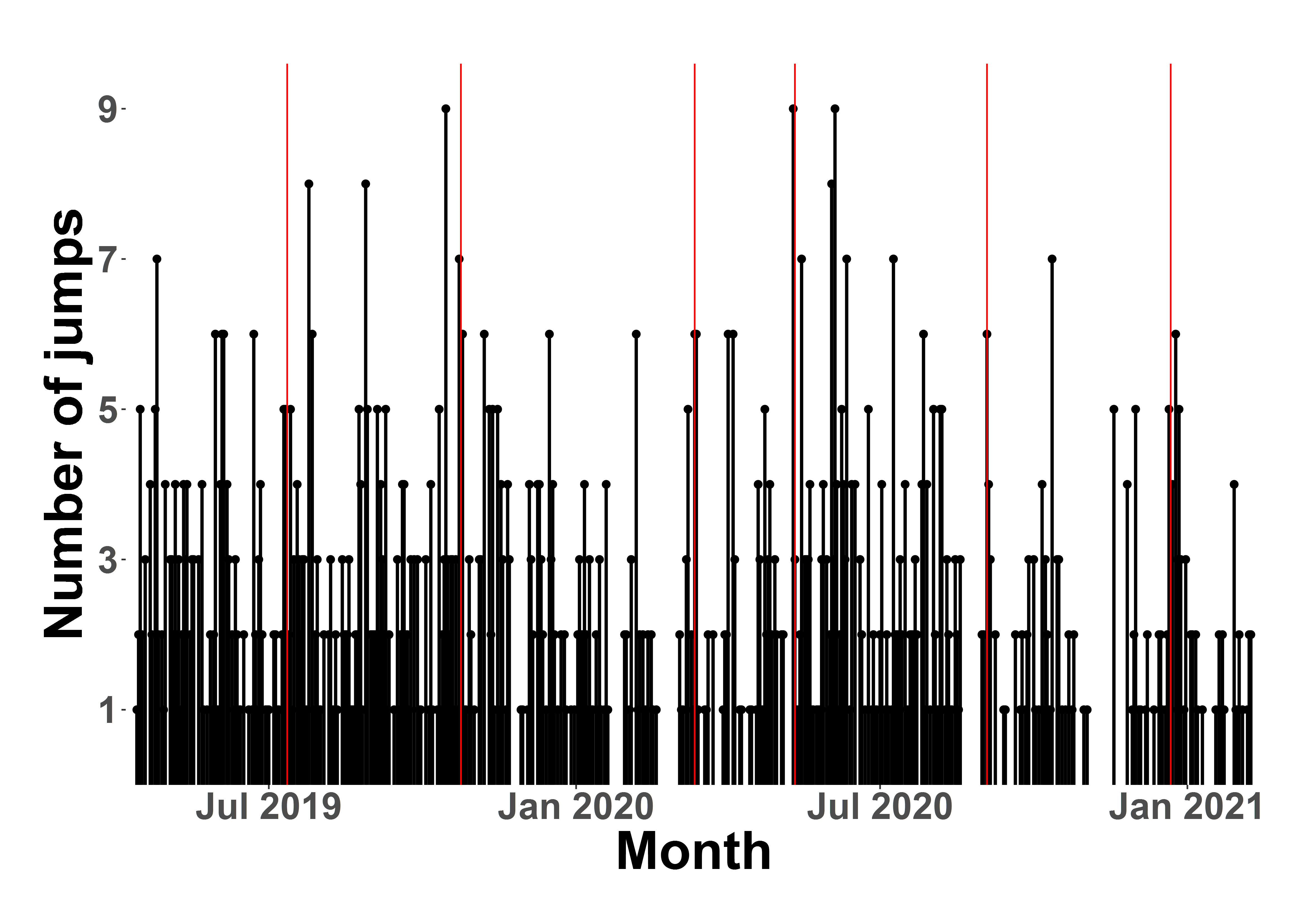}
    \caption{\label{fig:jumps_lm} Number of observed jumps on all time series. \protect \includegraphics[height=0.5cm]{plots/qletlogo_tr.png} {\color{blue}\href{https://github.com/QuantLet/JumpDetectR}{JumpDetectR}}}
\end{figure}
\FloatBarrier

Figure \ref{fig:jumps_lm} shows the number of jumps detected per day in all time series. We see that the number of jumps is varying over time, while prices and liquidity have largely increased both in comparison to the period of 2011-13 and 2019 vs 2021. The red vertical lines mark the same events as in figure \ref{fig:btctimechange}. This not exhaustive list of events includes a series of Trump tweets (July 2019), bullish comments of Xi Jinping (October 2019), the Covid crisis (March 2020), the third BTC halving (May 2020), large transfers to exchange wallets (September 2020), and the lawsuit against Ripple (December 2020). While we focused on only a few key events in the crypto universe, they seem to have an influence both on prices and on the number of jumps detected. Even though jumps are frequently occurring, they seem to be clustered in the neighborhood of these events, and possibly other events. A higher frequency of jumps often spreads over several days. Looking at the nature of price movements around these events, this finding is not surprising as these events were often preceded and/or followed by increased volatility.

Table \ref{jumps_days} shows the number of test days and the number of jumps detected. We observe that more testing days do not necessarily cause more jumps. This indicates that spurious jumps are not dominating the results. Moreover, larger assets tend to jump more often, and the the most jump were detected in the largest asset BTC. Most CCs tend to have more jump days than traditional assets, similarly to \textcite{scaillet_high-frequency_2020} who find that BTC has an unusually high jump rate.

\begin{table}[ht] \centering 
\begin{tabular}{lrrr} 
\\[-1.8ex]\hline 
\hline \\[-1.8ex] 
Symbol & N jumps & N test days & $\%$ jumps \\ 
\hline \\ 
BCH & $37$ & $137$ & $27.01$ \\ 
LTC & $51$ & $169$ & $30.18$ \\ 
ETC & $53$ & $110$ & $48.18$ \\ 
XRP & $158$ & $434$ & $36.41$ \\ 
ETH & $324$ & $559$ & $57.96$ \\ 
BTC & $423$ & $645$ & $65.58$ \\ 
\\[-1.8ex]\hline 
\hline \\[-1.8ex] 
\end{tabular} 
\caption{  \label{jumps_days} Number of test days and jumps per asset.} 
\end{table} 

Table \ref{sum_km} shows values for $k$ and $M$, and the block size for jump detection $k*M$ when applying the methodology of Lee \& Mykland. The block sizes tend to be quite high. Therefore, the exact moment of jump detection can only be approximated. Even though we can sample data at highest frequency, we often have to sub-sample to eradicate market microstructure noise. This is due to high values in the ACF test which serves as an indicator for dependent noise. In the original paper, the authors determine $k=3$ empirically, whereas we often have to choose values of $k>10$. The theoretical properties of existing jump detection methods are not fully applicable to CCs. These findings call for new methods that can properly capture the distinct market microstructure of digital asset markets.

\begin{table}[ht]
\centering 
\begin{tabular}{lrrrrrrr} 
\\[-1.8ex]\hline 
\hline \\[-1.8ex] 
Statistic & \multicolumn{1}{r}{Mean} & \multicolumn{1}{r}{St. Dev.} & \multicolumn{1}{r}{Min} & \multicolumn{1}{r}{Pctl(25)} & \multicolumn{1}{r}{Median} & \multicolumn{1}{r}{Pctl(75)} & \multicolumn{1}{r}{Max} \\ 
\hline \\
$k$ & 13.529 & 14.436 & 3 & 5 & 5 & 17 & 51 \\ 
$M$ & 4.235 & 1.765 & 1 & 3 & 4 & 6 & 11 \\ 
$kM$ & 43.389 & 29.158 & 15 & 25 & 30 & 52 & 194 \\ 
\\[-1.8ex]\hline 
\hline \\[-1.8ex] 
\end{tabular} 
  \caption{  \label{sum_km}  Summary statistics of $k$ \& $M$ (rounded).} 
\end{table}

\begin{figure}[ht]
\centering
\begin{subfigure}{.4\textwidth}
\includegraphics[width=\textwidth]{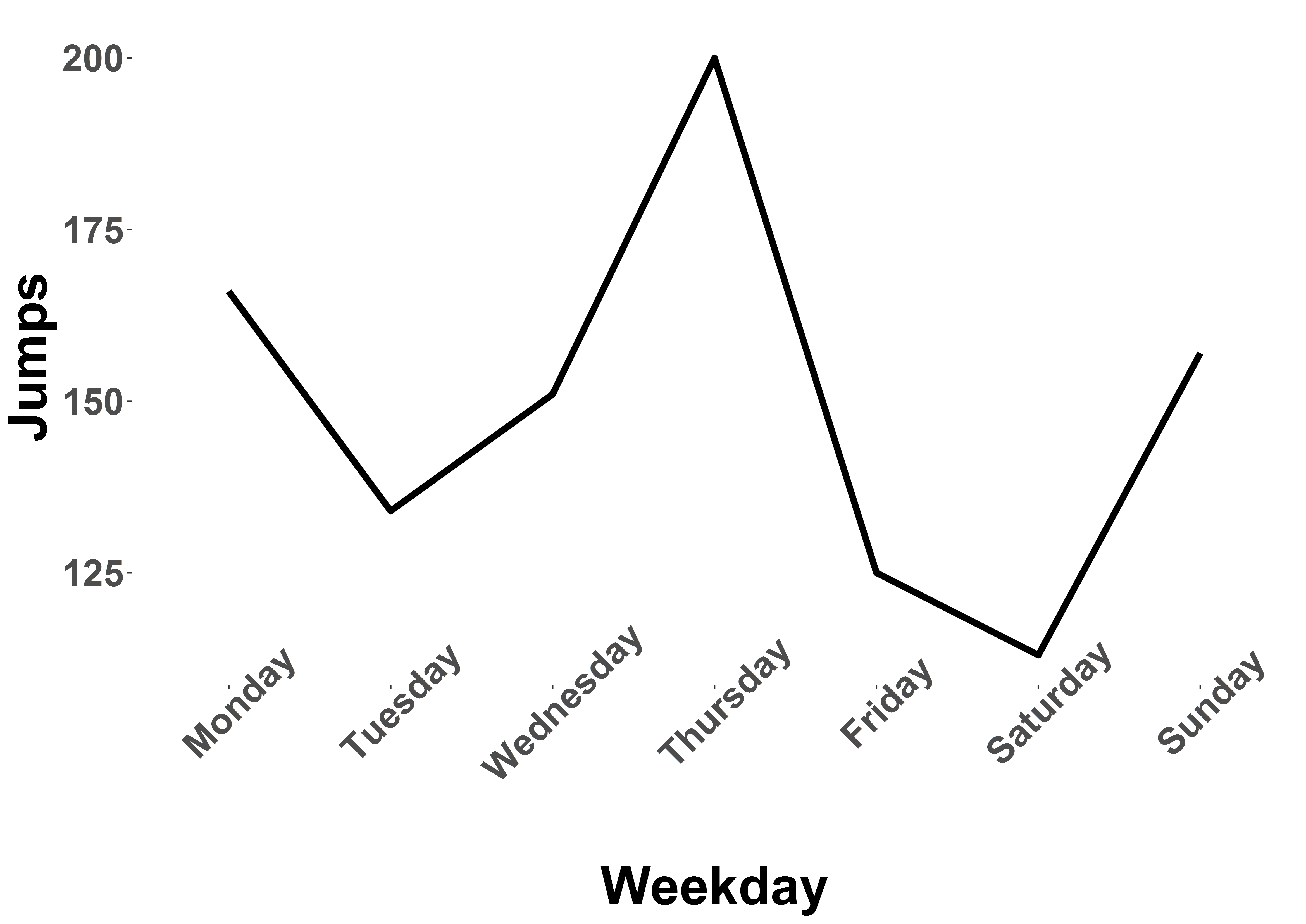}
\end{subfigure}
\begin{subfigure}{.4\textwidth}
\includegraphics[width=\textwidth]{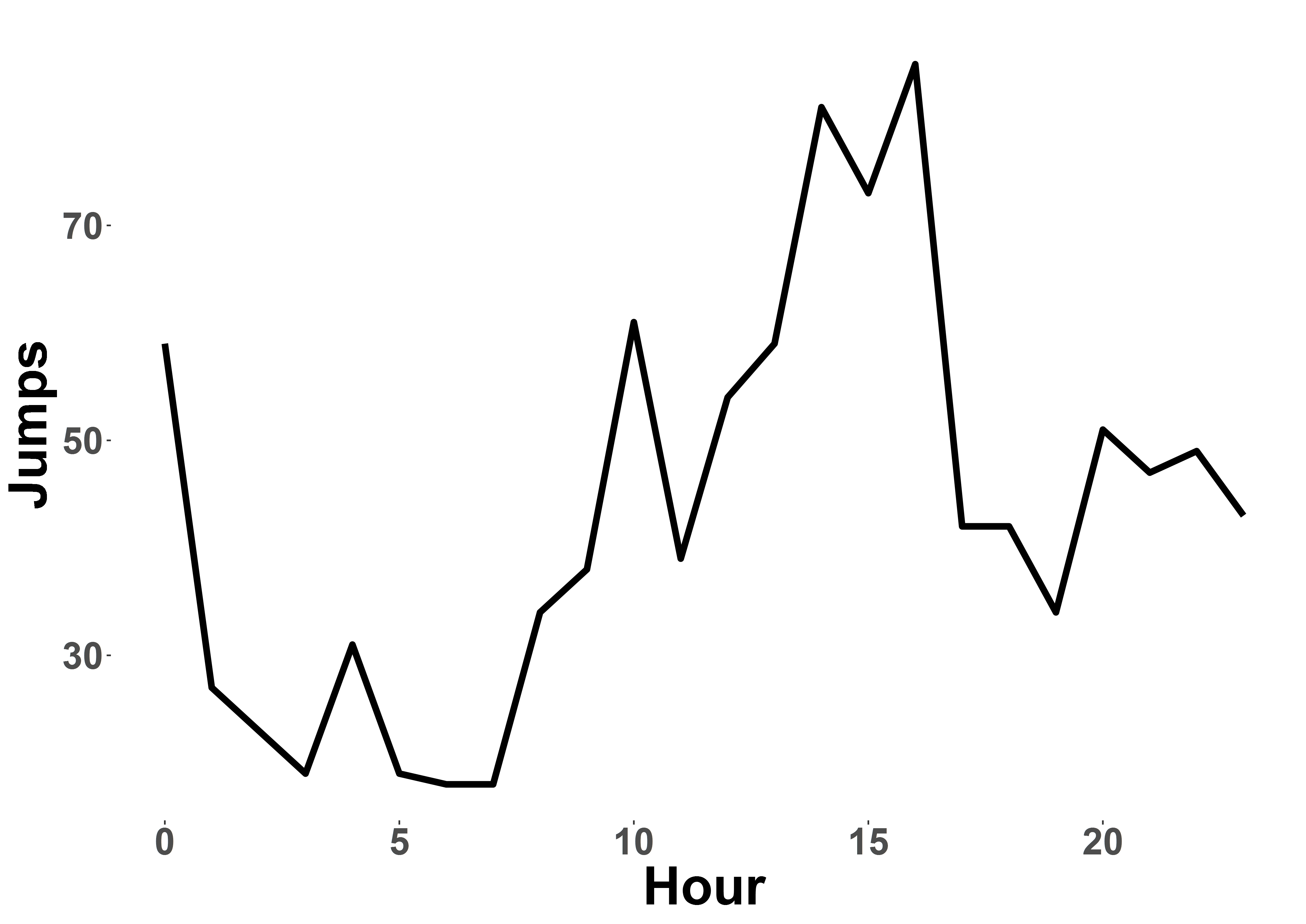}
\end{subfigure}
\\
\caption{Number of jumps per weekday and per hour (both aggregated). \protect \includegraphics[height=0.5cm]{plots/qletlogo_tr.png} {\color{blue}\href{https://github.com/QuantLet/JumpDetectR}{JumpDetectR}}
}
\label{fig:jumps_by_day_h}
\end{figure}

Figure \ref{fig:jumps_by_day_h} shows the number of jumps aggregated per weekday and per hour. Looking at weekdays, the highest number of jumps is observed during the middle of the week, whereas the lowest values on are on Friday and Saturday. This is an interesting finding, as \textcite{petukhina_rise_2021} shows similar patterns in volatility and trading volume on different weekdays. We see a strong connection between volatility and trading volume, and the occurrence of jumps. Additionally, we find evidence for seasonality patterns in the detection of jumps per hour, as most jumps are detected around 13-17h UTC and the lowest number of jumps between 1-7h UTC. This is an interesting finding as the time window of 13-17h UTC is the one where it is plausible for people ranging from the US, to Europe, and East Asia to be awake (assuming that we could expect the majority of people to be awake between 8am and 10pm).

\FloatBarrier

\begin{figure}[ht]
    \centering
    \includegraphics[width=0.6\textwidth]{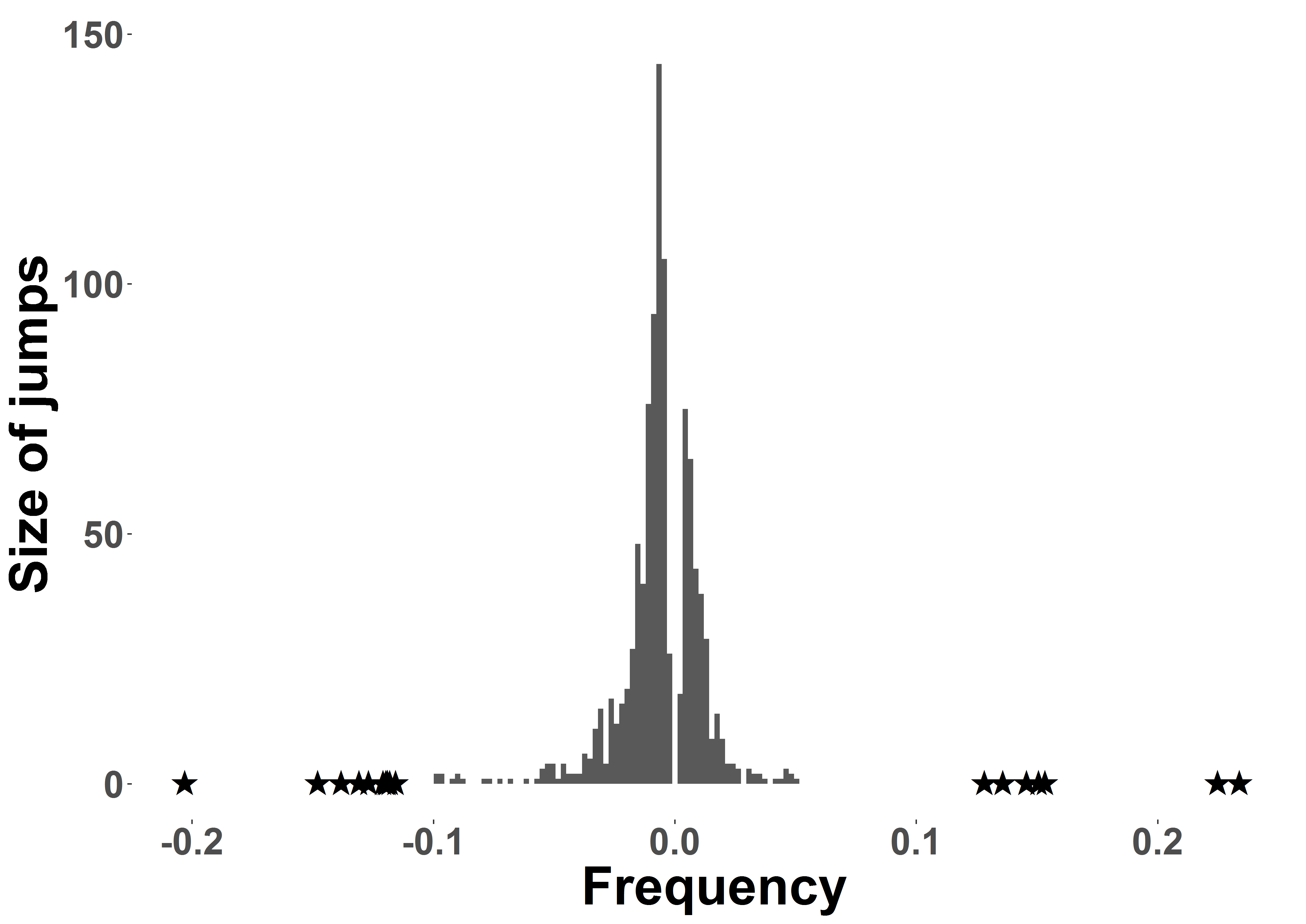}
    \caption{Size of jumps. \protect \includegraphics[height=0.5cm]{plots/qletlogo_tr.png} {\color{blue}\href{https://github.com/QuantLet/JumpDetectR}{JumpDetectR}}}
    \label{fig:jumps_hist}
\end{figure}

\begin{table}[ht]
  \begin{subtable}{\linewidth}
  \centering
    \begin{tabular}{lrrr}
        \\[-1.8ex]\hline 
        \hline \\[-1.8ex] 
        Statistic & All & Positive & Negative \\ 
          \hline
        N & 1,046 & 334 & 712 \\ 
         Min. & -0.203 & 0.001 & -0.203 \\ 
          1st Qu. & -0.012 & 0.005 & -0.017 \\ 
          Median & -0.006 & 0.008 & -0.010 \\ 
          Mean & -0.006 & 0.014 & -0.016 \\ 
          3rd Qu. & 0.005 & 0.013 & -0.006 \\ 
        Max. & 0.234 & 0.234 & -0.002 \\ 
          Skewness & 0.644 & 6.211 & -4.160 \\ 
          Kurtosis & 27.757 & 46.529 & 25.025 \\ 
        \\[-1.8ex]\hline 
        \hline \\[-1.8ex] 
  \end{tabular}
  \caption{Summary statistics of jump size.} 
  \label{sum_jumps} 
  \end{subtable}
    \begin{subtable}{\linewidth}
  \centering
  \begin{tabular}{@{\extracolsep{5pt}} lclc} 
\\[-1.8ex]\hline 
\hline \\[-1.8ex] 
Negative & Counts & Positive & Counts  \\ 
\hline \\[-1.8ex] 
$< -0.025$ & $108$ & $> 0.025$ & $27$ \\
$< -0.05$ & $33$ & $> 0.05$ & $8$ \\
$< -0.1$ & $11$ & $> 0.1$ & $7$ \\ 
$< -0.2$ & $1$ & $> 0.2$ & $2$  \\ 
\\[-1.8ex]\hline 
\hline \\[-1.8ex] 
\end{tabular} 
  \caption{Extreme jumps.} 
  \label{extreme_jumps} 
  \end{subtable}
\caption{Descriptive statistics on jumps.} 
\end{table}

Figure \ref{fig:jumps_hist} presents the histogram of jump sizes. It shows a large density of jumps in the area of +-5\% and especially in the area of +-2.5\%, and a dominance of negative jumps. Furthermore, the distribution has heavy tails in both directions. This is a common property for asset returns as well, and provides further evidence for a relation between jumps and returns. Table \ref{sum_jumps} shows the according summary statistics on jumps, aggregated and separated by positive and negative jumps. Most jumps are rather small, and despite the fact that roughly $2/3$ of jumps are negative, the overall distribution of jumps is positively skewed and the positive tail is heavier than the negative one. This is due to the selected time frame in which the Covid crisis falls, but also a recovery in a later stage with large positive returns.

In contrast to the common assumption that jumps happen mainly during flash crashes, the results indicate that prices do not seem to jump heavily during bearish periods. It seems that crash patterns resemble a quick but steady sell-off, rather than concentrated, heavy movements. Bullish periods with large upwards price movements seem to create heavier jumps on the other hand, such that these movements resemble rallies where investors rapidly join and create price jumps. While CCs undoubtedly have a large downside risk due to their large volatility, this risk was outweighed by the high rewards of the latest bull run. At least in the observed time frame, ETFs and investors may have largely profited from these characteristics of CCs. In addition to the many small jumps, few large jumps can be observed in the tails, as can be seen in table \ref{extreme_jumps}. Even though most jumps are in the area of +-$1\%$, we can observe more than 100 returns smaller than -2.5\% and 27 returns larger than +2.5\%. 33 observed HF returns are smaller than -5\% and 8 returns are larger than +5\%. Looking at jumps with magnitude of +-10\%, we still observe 7 (11) jumps, whereas 2 (1) returns are +-20\%. This is a result of the preprocessing done in the Lee \& Mykland / Aït-Sahalia, Jacod and Li jump methodology w.r.t. previously discussed challenges when working with HF data. While this clearly blurs the exact moment of an occurred jump, it also means that CCs are prone to jump by more than +-20\% within few minutes or seconds during strongly bullish / bearish phases. Even though extreme jumps are much rarer than extreme returns, they still seem to be a reoccurring phenomenon. And despite the obvious differences, the statistical properties of extreme returns and actually detected jumps look similar. Therefore, we now turn to the relationship between jumps and returns. In comparison with \textcite{scaillet_high-frequency_2020}, we find that with the time being, the skewness of the jump distribution, while still positive, has decreased significantly over all time series.

Table \ref{table:reg_jumps} presents the results of the regression of daily returns against the occurrence of a jump on the same day, the occurence of a jump on the previous day, and the occurence of a positive (negative) jump on the same day. The regression tables show the slope estimates and the corresponding standard deviations in brackets below. We report significant observations with three stars for a $p$-value $< 0.001$, two stars for $p <0.01$, and one star for $p < 0.05$.  To guarantee robust estimates, we use a one-way fixed effects estimator with a within transformation to account for the differences between the different CCs. We use White standard errors for heteroscedasticity-consistency. We find that jumps significantly affect end of day returns. Given the dominance of negative jumps in the dataset, the overall effect is negative on daily returns. Accordingly, when separating between positive and negative jumps, jumps either have a positive or negative effect on the return process. A jump on the previous day does not significantly affect daily returns in our dataset. The analysis shows that intra-day detection of jumps has implications on end-of-day returns in the same direction as a detected jump, however not on the following day. Since daily returns in our dataset were negatively skewed for all CCs and the vast majority of detected jumps was also negative, the skewness of a distribution seems to be related to the ratio of positive vs. negative jumps. However, judging from table \ref{sum_jumps} this applies only to the number of jumps, but not to their respective size, as the jump size distribution is positively skewed despite having observed mostly negative jumps.

\begin{table}[ht]
\begin{center}
\begin{tabular}{l D{.}{.}{4.5} D{.}{.}{4.4} D{.}{.}{4.4} D{.}{.}{4.6}}
\hline
\hline
 & \multicolumn{1}{c}{Jumps (all)} & \multicolumn{1}{c}{Lagged jumps (all)} & \multicolumn{1}{c}{Jumps (pos.)} & \multicolumn{1}{c}{Jumps (neg.)} \\
\hline
jump\_dummy      & -0.017^{**} & 0.001   &           &              \\
                 & (0.006)     & (0.003) &           &              \\
pos\_jump\_dummy &             &         & 0.014^{*} &              \\
                 &             &         & (0.006)   &              \\
neg\_jump\_dummy &             &         &           & -0.029^{***} \\
                 &             &         &           & (0.005)      \\
\hline
R$^2$            & 0.016       & 0.000   & 0.004     & 0.037        \\
Adj. R$^2$       & 0.013       & -0.003  & 0.002     & 0.035        \\
Num. obs.        & 2040        & 2040    & 2040      & 2040         \\
\hline
\hline
\multicolumn{5}{l}{\scriptsize{$^{***}p<0.001$; $^{**}p<0.01$; $^{*}p<0.05$}}
\end{tabular}
\caption{Relationship of jumps and daily returns.}
\label{table:reg_jumps}
\end{center}
\end{table}
\FloatBarrier

\section{Conclusion} \label{sec:conclusion}
Previous research did not incorporate cross-market data and several of the largest CCs. Jumps are an essential component in the price process not only of BTC, but also of other CCs. In addition, the frequency of jumps has not decreased, and maybe even increased, however this is hard to evaluate due to different sampling frequencies. While we investigate jumps in HF, previous analyses sampled at 5 minutes intervals. Jumps in both directions siginficantly affect end-of-day returns. Additionally, jump effects do not seem to be persistent, i.e. they only affect returns on the same day. However, given that the size of daily returns is larger than the detected jump sizes, jumps seem to explain the direction and intensity of returns. Moreover, since the CC universe has grown, researchers should increasingly shift their focus to events at many exchanges at once and investigate also how jumps are correlated between CCs. Furthermore, the robustness of methodologies for CCs should be evaluated, as e.g. the limiting behavior of the daily variation is not similar to traditional to that of assets in \textcite{lee_jumps_2012}. While \textcite{ait-sahalia_testing_2012} seems more robust to the differing variation, it provides no information on the moment of jump detection. Possible future work could extend these methodologies to provide robust intra-day information on the occurrence of jumps. It would be interesting to investigate whether there are patterns in the order of jump detection order, e.g. if jumps in certain currencies most likely also trigger jumps in other currencies, and if so, when. Studying the effects of these jumps on the system of currencies could allow us to investigate contagion dynamics and thus improve our understanding of the CC universe. The relationship between volatility and trading volume, and the occurrence of jumps is another interesting direction that future research could follow. The few events in the CC space that we studied revealed that they clearly relate to the occurence of jumps. Future studies should focus on a more systematic approach on manifesting this relationship, since jumps clearly need to be incorporated in any meaningful option pricing model. 
\clearpage
\singlespacing
\printbibliography
\clearpage
\section*{Appendix A. Lee \& Mykland jump test} \label{sec:app_lmjumptest}
\addcontentsline{toc}{section}{Appendix A}
Due to market microstructure noise, it is impossible for us to observe the true price, we can only observe
\begin{equation*}
\tilde{P_{t}} = X_{t} + \epsilon_t,
\end{equation*}
the price contamined with noise $\epsilon_t \subset \mathbb {R} $ with  standard deviation $q \in \mathbb {R}^+ $. Denote as $\hat\sigma \in \mathbb {R}^+ $ the volatility estimate, which we calculate. We denote $k-1$, where $k \in \mathbb {N}$, as the lag order of the autocorrelation function of $\tilde{P_{t}}$ that is determined empirically and fix the grid
\begin{equation*}
\mathcal{G}^k_n = \left \{ 0 = t_{n,0} < t_{n,k}  < t_{n,2k} < \cdots  \right \} \end{equation*}
to get the subsampled price $\tilde{P}(t_{ik})$ consisting of only observations inside $\mathcal{G}^n_k$ and introduce
\begin{equation*}
\mathcal{G}_n^{kM} = \left \{ t_n < t_{n,kM} <  t_{n,2kM} < \cdots \right \} = \left \{ t_0 < t_{kM} <  t_{2kM} < \cdots \right \} 
\end{equation*}
in order to sample $\hat{P}(t_j)$ at every $M$ observations from $\mathcal{G}^k_n$ with $t_j \in \mathcal{G}_n^{kM}$ for all $j$ and $M \sim C\left \lfloor n/k \right \rfloor^{\frac{1}{2}}$. $C \in \mathbb {R}^+$ is selected as per recommendation of \textcite{lee_jumps_2012} and \textcite{jacod_microstructure_2009, jacod_limit_2010}. Now, define 
\begin{equation*}
\hat{P}(t_j) \overset{\text{def}}{=}  M^{-1}\sum_{i = \left \lfloor j/k\right \rfloor}^{\left \lfloor j/k\right \rfloor + M-1}\tilde{P}(t_{ik})
\end{equation*} 
and calculate the returns of these pre-averaged prices
\begin{equation*}
\bar{P}(t_j) \overset{\text{def}}{=}  {\hat{P}(t_{j+kM})- \hat{P}(t_{j})}
\end{equation*}
and scale these returns 
\begin{equation*}
\mathcal{\chi}(t_j) \overset{\text{def}}{=} { \frac{\sqrt{M}}{\sqrt{V_n}} \bar{P}(t_j)}
\end{equation*}
with $\mathcal{\chi}(t_j)$ following a standard normal distribution, and
\begin{equation*}
V_n \overset{\text{def}}{=}  {\mbox{Var}\left [ \sqrt{M}\bar{P}(t_j)\right ]}. \label{eq:Vn}
\end{equation*} 
Note that $V_n$ has the limit 
\begin{equation*}
\plim_{n \to \infty }V_n = \frac{2}{3}\sigma^2C^2T+2q^2, \label{eq:plim}
\end{equation*} 
where 
\begin{equation*}
\hat{q}^2=\frac{1}{2(n-k)}\sum^{n-k}_{m=1}\{\tilde{P}(t_{m})-\tilde{P}(t_{m+k})\}^2 \label{eq:qhat}
\end{equation*}
and $\hat\sigma$ are used to estimate noise variance and volatility respectively, s.t. we can calculate the test statistic
\begin{equation*}
\hat\xi_{t_j} \overset{\text{def}}{=}  {\frac{| \chi\left ( t_j \right ) | - A_n}{B_n}}
\end{equation*}
for every $\bar{P}_{t_j}$. Using the result that
\begin{equation*}
\frac{\max_{t_j\in\mathcal{G}_n^{kM}}| \chi\left ( t_j \right ) | - A_n}{B_n}\overset{\mathcal{L}}{\rightarrow}\xi
\end{equation*}
asymptotically, where $\xi$ follows a standard Gumbel distribution,  with the scaling terms
\begin{equation*}
A_n = \left ( 2 \log\left \lfloor \frac{n}{kM} \right \rfloor \right )^{1/2} - \frac{\log\pi+\log\left ( \log\left \lfloor \frac{n}{kM} \right \rfloor \right )}{2\left (2\log\left \lfloor \frac{n}{kM} \right \rfloor  \right )^{1/2}}, 
\end{equation*}
\begin{equation*}
B_n = \frac{1}{\left (2\log\left \lfloor \frac{n}{kM} \right \rfloor  \right )^{1/2}},
\end{equation*}
under the null hypothesis of no jumps we can then say that if e.g. $\hat \xi >$ 99th percentile of the standard Gumbel distribution we observe a jump.

\section*{Appendix B. Aït-Sahalia, Jacod and Li jump test} \label{sec:app_ajljumptest}
\addcontentsline{toc}{section}{Appendix B}
Recall the results of \textcite{ait-sahalia_testing_2009}. We have $n$ observed increments of $Y$ on $[0,t]$. For any integer $i \geq 1$ and $p \in \mathbb {R}^+$ we can write 
\begin{equation}
    \Delta _{i}^{n}Y = Y_{i\Delta_n } - Y_{(i-1)\Delta_n},\text{   } B (Y,p,\Delta_n)_t = \sum_{i=1}^{[t/\Delta_n]}|\Delta_i^n|^{p}. \label{eq:B_p}
\end{equation}
For an integer $k \geq 2$ the test statistic is denoted as
\begin{equation*}
    S_J ( p,k,\Delta_n)_n = \frac{B(X,p ,k\Delta_n)_T}{B(X,p ,\Delta_n)_T}.\label{eq:S_J}
\end{equation*}
Without noise, this can be calculated from the data. There are two possible tests where the set $\Omega_T^j$ means that jumps are observed and $\Omega_T^c$ that a continuous path is observed. For $p>2$, they have the asymptotic behavior
\begin{equation*}
S_J \left ( p,k,\Delta_n \right ) \overset{\mathbb{P}}\rightarrow \left\{\begin{matrix}
1 & \text{on } \Omega_T^j  \\ 
k^{p/2-1} & \text{on } \Omega_T^c  
\end{matrix}\right.
\end{equation*}
Since this test statistic is not robust to noise, the construction of a robustified test statistic is necessary. This task makes again use of the results from \textcite{jacod_microstructure_2009, jacod_limit_2010}.
For defining a pre-averaging window, a sequence of integers $k_n$ needs to be chosen that satisfies:

\begin{equation*}
    k_n\sqrt{\Delta_n} = \theta + { \scriptstyle \mathcal{O}}(\Delta_n^{1/4}), \theta > 0.
\end{equation*}
Weight functions $g \in \mathbb{R}$ are used to weigh observations within the pre-averaging window, where
\begin{equation*}
    \left.\begin{matrix}
g \text{ is continuous, piece-wise } C^1 \\ \text{    with a Lipschitz derivative }g', \\ 
s  \not\in (0,1) \Rightarrow g(s) =0, & \text{   }\displaystyle \int g(s)^2ds >0,
\end{matrix}\right\},
\end{equation*}
with which we can obtain the parameters
\begin{equation*}
    \left.\begin{matrix}
g_i^n = g(i/k_n), & g_i^{'n}=g_i^n-g_{i-1}^n, \\ 
\bar{g}(p) = \displaystyle \int \abs{g(s)}^{p} ds, & \bar{g}^{'}(p) = \displaystyle \int \abs{g(s)^{'}}^{p} ds.
\end{matrix}\right\}.
\end{equation*}

Now, for any $Y = (Y_t)_{t \geq 0}$ we have the random variables

\begin{equation*}
    \bar{Y}(g)_i^{n}=\sum_{j=1}^{k_n -1}g_j^{n}\Delta_{i+j}^{n}Y, \text{   } \hat{Y}(g)_i^{n}=\sum_{j=1}^{k_n}(g_j^{'n}\Delta_{i+j}^{n}Y)^2
\end{equation*}
as well as the processes
\begin{equation*}
    V(Y,g,q,r)_t^n = \sum_{i=0}^{[t/\Delta_n]-k_n}\abs{\bar{Y}(g)_i^{n}}^{q}\abs{\hat{Y}(g)_i^{n}}^r,
\end{equation*}
and these processes implicitly depend on $\Delta_n$ and $k_n$. 

Let $p\geq4$ and even integer, define $(\rho(p)_j)_{j=0,...,p/2}$ as the unique numbers solving the triangular system of linear equations 
\begin{equation*}
  \left.\begin{matrix}
\rho(p)_0=1, \\ 
\sum_{l=0}^{j}2^{l}m_{2j-2l}C_{p-2l}^{p-2j}\rho(p)_{l}=0, j = 1,2,...,p/2,
\end{matrix}\right\},    
\end{equation*}
with $m_r$ the $r$th absolute moment of the law $\mathcal{N}(0,1)$. We follow the recommendation of the authors and set $p=4$, s.t. we obtain
\begin{equation*}
\rho(4)_0=1, \text{ }  \rho(4)_1=-3, \text{ }  \rho(4)_2=0.75, 
\end{equation*}
fix $k_n =100$ and for any process $Y$ we set
\begin{equation*}
\bar{V}(Y,g,p)_t^n = \sum_{l=0}^{p/2}\rho(p)_{l}V(Y,g,p-2l,l)_t^n,
\end{equation*}
which is a robustified version of the power variation in \ref{eq:B_p}.

To compute the robustified test statistic for jumps we set up the constants

\begin{equation*}
    \gamma = \frac{(\bar g)(2)}{(\bar h)(2)},  \gamma^{'} = \frac{(\bar g)(p)}{(\bar h)(p)},  \gamma^{''} = \frac{\gamma^{p/2}}{\gamma^{'}},
\end{equation*}
under the assumption that $\gamma^{''} >1$. The robustified test statistic is
\begin{equation*}
    S_{RJ}(g,h,p)_n = \frac{\bar{V}(Z,g,p)_T^n}{\bar{V}(Z,h,p)_T^n}.\label{eq:S_RJ}
\end{equation*}

Asymptotically, the test statistic has the following limit behavior:
\begin{equation*}
S_{RJ} \left ( p,k,\Delta_n \right ) \overset{\mathbb{P}}\rightarrow \left\{\begin{matrix}
1 & \text{on } \Omega_T^j  \\ 
\gamma^{''} & \text{on } \Omega_T^c  
\end{matrix}\right.
\end{equation*}

Introduce the variance scaling term 
\begin{equation*}
    \sqrt{\Sigma_{RJ,n}^c} = \frac{M*(g,g,\phi;p)_T^n-2\gamma^{p/2}M*(g,h,\phi,p)_T^n+\gamma^{p}M*(h,h,\phi;p)_T^n}{(\Delta_n^{1-p/4}\bar{V}(Z,g,p)_T^{n}/\gamma^{''})^2},
\end{equation*}
then the critical value for rejecting the null hypothesis of no jumps is 
\begin{equation*}
C_n^c =  \Bigl\{S_{RJ}(g,h,p)_n <  \gamma^{''}- z_{\alpha} \Delta_n^{1/4} \sqrt{\Sigma_{RJ,n}^c} \Bigr\}.
\end{equation*}
with $ z_{\alpha}$ the corresponding quantile of the standard normal distribution, s.t. the rejection region can be obtained by choosing a respective $\alpha$.

\end{document}